\pdfoutput=1
\documentclass[12pt]{iopart}
\usepackage[utf8]{inputenc}
\usepackage{iopams}
\usepackage{amsfonts}
\usepackage{graphicx}

\newcommand{\muG}{$\mu$Gal$\;$}
\newcommand{\D}{\mathrm{d}}
\newcommand{\w}{\omega T /2}

\begin{document}
\maketitle
\title{Enabling a linear model for the IMGC-02 absolute gravimeter}
\author{V D Nagornyi}
\address{Metromatix, Inc., 111B Baden Pl, Staten Island, NY 10306, USA}
\ead{vn2@member.ams.org}
\author{E Biolcati}
\address{Istituto Nazionale di Geofisica e Vulcanologia (INGV), Sede di Portovenere, Fezzano (SP), Italy}
\author{S Svitlov}
\address{Institute of Optics, Information and Photonics Friedrich-Alexander University Erlangen-Nuremberg, Germany}

\begin{abstract}
Measurement procedures of most rise-and-fall absolute gravimeters have to resolve singularity at the apex of the trajectory caused by the discrete fringe counting in the Michelson-type interferometers.
Traditionally the singularity is addressed by implementing non-linear models of the trajectory, but they introduce problems of their own, such as biasness, non-uniqueness, and instability of the gravity estimates.
Using IMGC-02 gravimeter as example, we show that the measurement procedure of the rise-and-fall gravimeters can be based on the linear models which successfully resolve the singularity and provide rigorous estimates of the gravity value.
The linear models also facilitate further enhancements of the instrument, such as accounting for new types of disturbances and active compensation for the vibrations.
\end{abstract}
\newpage
\section{Introduction}
Absolute ballistic gravimeters measure gravity acceleration by tracking the free motion of the test mass in the gravity field. Using the positions $\{S_1, ..., S_N\}$ of the test mass at the moments  $\{T_1, ..., T_N\}$, the acceleration is found as parameter of some trajectory model $z(t)$ fitted to the data pairs $(T_i, S_i)$.
There are two ballistic techniques known in absolute gravimetry: direct free-fall and symmetric rise-and-fall ones. In the free-fall gravimeters the test mass is released from its upper position and tracked during the free fall to its lower position. In rise-and-fall gravimeters the test mass is thrown up vertically and tracked on both upward and downward parts of the trajectory.
Realization of each technique leads to different designs and uncertainty budgets of the instruments.
The importance of this diversity for gravimetric metrology was highlighted at the very first International Comparison of Absolute Gravimeters \cite{boulanger1983}. Analyzing the result of the Italian instrument, M.~U.~Sagitov found \cite{sagitov1984} that the lower reported value was caused by the higher order terms in the vertical gravity gradient influencing the result through the gravimeter's lower effective position. Since then the vertical variations of the gradient are routinely considered in deriving comparison reference values. 
Theoretical analysis shows that the rise-and-fall instruments can achieve the same or even better uncertainty compared to the free-fall ones \cite{faller1988}. Still, symmetric instruments are very few in the world and rarely participate in the comparisons. This fact can be explained by several specific problems inherent to this type of instruments. The main problem is more difficult mechanical realization of the launching device of the rise-and-fall gravimeters, compared to a more simple release device of the free-fall instruments. Another major problem is related to the singularity at the apex of the symmetric trajectory. The Michelson-type interferometers commonly used in absolute gravimeters to track the test mass are insensitive to the motion direction, so the downward branch of the trajectory gets reflected upwards creating a jump in the acceleration of about $2g$. Traditionally the singularity was resolved by implementing non-linear models of the trajectory to derive the gravimeter measurement equation. In this paper we develop a different approach to the problem producing the measurement equations based only on linear models of the trajectory.
We then compare two approaches implemented for the IMGC-02 gravimeter and conclude the advantages of the linear model in the context of rise-and-fall gravimeters.
\section{Apex singularity}
\label{sec_apex_singularity}
\begin{figure}[t]
\centering
\includegraphics[height=100mm]{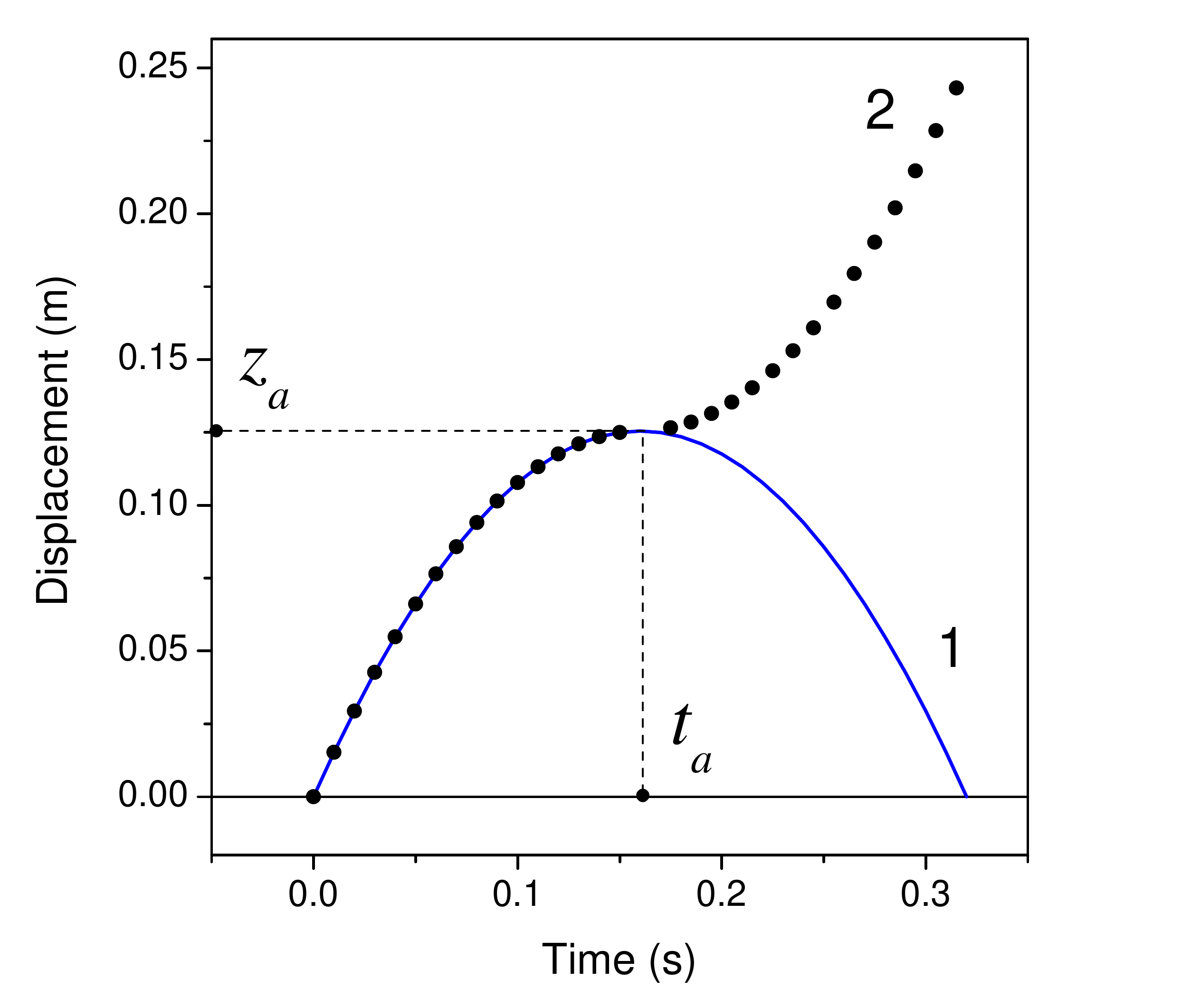}
  \caption[short title]
  {
  Symmetric rise-and-fall trajectory:
  1 -- actual;
  2 -- as tracked by laser (Michelson-type) interferometer.
  }
\label{fig_2_traj}
\end{figure}

Every period of the fringe signal of the gravimeter interferometer corresponds to the test mass advancing one-half of the laser wavelength $\lambda/2$. The counting of the periods is incremental, so the registered trajectory looks like the test mass always moved upwards (fig.~\ref{fig_2_traj}). Due to this effect, even the simplest linear model of the trajectory describing undisturbed motion in the uniform gravity field
\begin{equation}
\label{eq_3_param}
z(t) = S_0 + V_0 t + g t^2/2
\end{equation}
turns into the much more complicated non-linear one (fig.~\ref{fig_2_traj})
\begin{equation}
\label{eq_2_traj}
z(t) = S_0-\frac{V_0^2}{2g}-
\left[ 
\textrm{sgn} \left( t+\frac {V_0}{2g} \right) 
\right] 
\left( V_0 t + g t^2/2 + \frac{V_0^2}{2g} \right)
\end{equation}

\begin{figure}[ht]
\centering
\includegraphics[height=100mm]{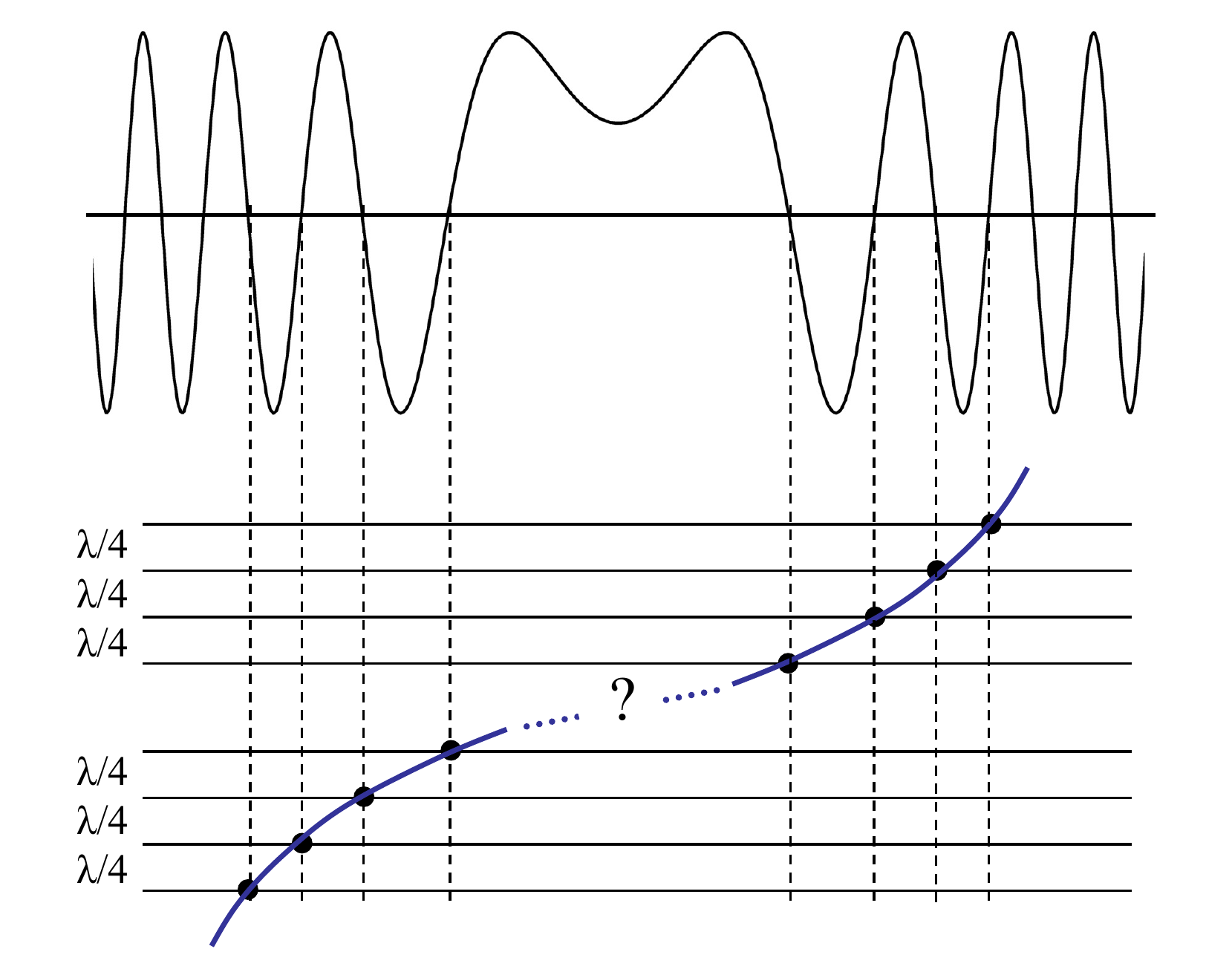}
  \caption[short title]
  {
  Uncertainty of the apex due to the interruption of the fringe count
     }
\label{fig_apex_fringe}
\end{figure}

\begin{figure}[t]
\centering
\small
\includegraphics[height=70mm]{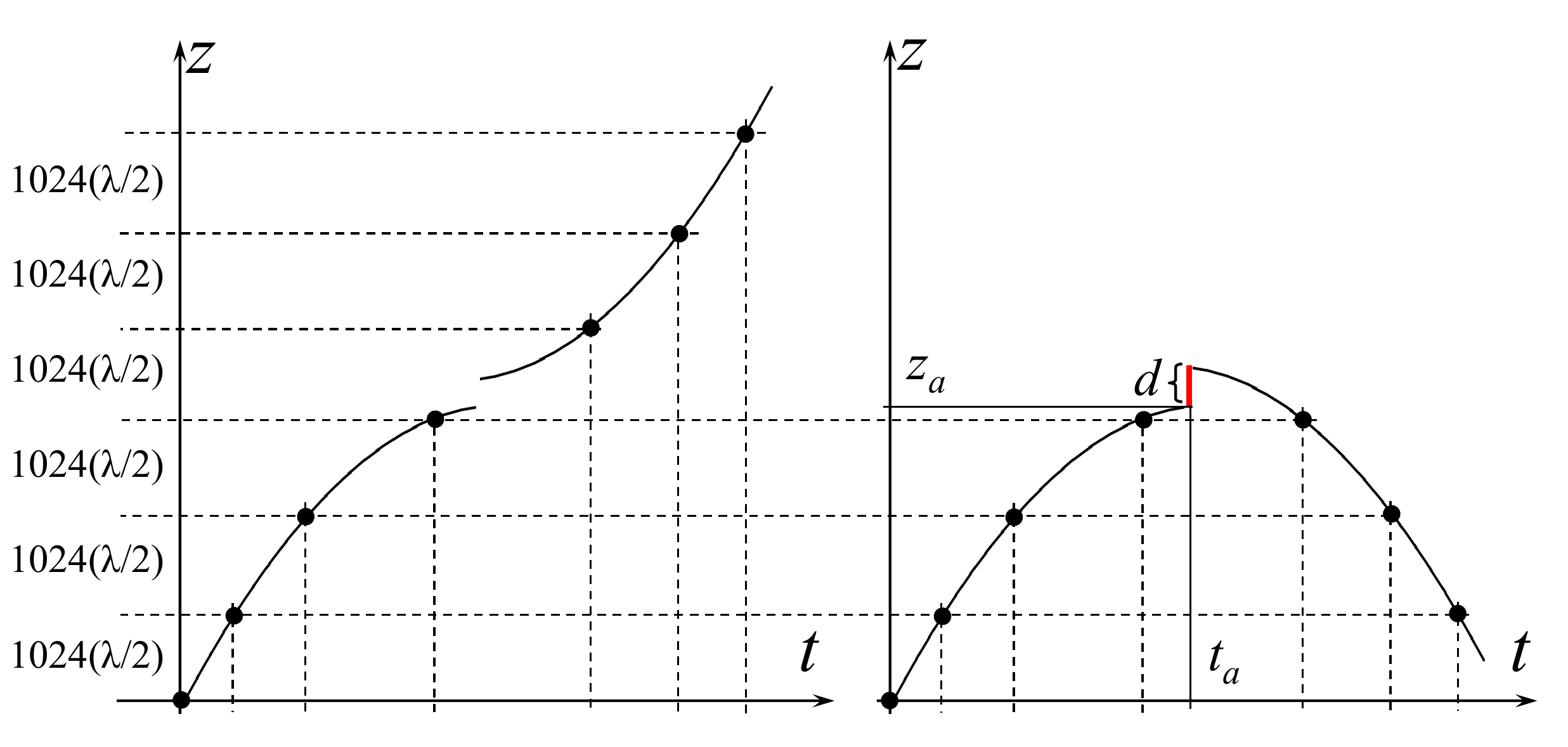}
  \caption[short title]
  {
Left -- registered trajectory,
right -- step in the recovered trajectory due to uncertainty of the apex coordinate
  }
\label{fig_apex_step}
\end{figure}

Let the apex coordinate be $(t_a, z_a)$. The actual trajectory ${\hat z}(t)$ is related to the registered one (\ref{eq_2_traj}) as
\begin{equation}
\label{eq_refl_trans}
{\hat z}(t) = \left\{
\begin{array}{ll}
z(t)   &   \,  t  < t_a, \\
z_a - z(t)  &   \,  t  \ge t_a,
\end{array}
\right.
\end{equation}
and can be written in the form
\begin{equation}
\label{eq_carre_smpl}
{\hat z}(t) = z_a + \frac{g}{2}(t - t_a)^2.
\end{equation}
The situation, however, is yet more complicated. The fringe signal does not necessarily cross zero at the apex, creating a gap in continuous counting of the periods (fig.~\ref{fig_apex_fringe}).
The gap gets much bigger in the IMGC-02 gravimeter, where the data are taken on every 1024-th fringe.  The gap adds an unknown displacement $d$ (fig.~\ref{fig_apex_step})
to the conversion (\ref{eq_carre_smpl}), so it becomes
\begin{equation}
\label{eq_carre_smpl_step}
{\hat z}(t) = z_a + \frac{g_a}{2}(t - t_a)^2 + d\,{\rm u}(t - t_a),
\end{equation}
where ${\rm u}(t)$ is the Heaviside step function:
\begin{equation}
\label{eq_Heaviside}
{\rm u}(t) = \left\{
\begin{array}{ll}
0 &  , \,  t< 0 \\
1  &  , \,  t \ge 0.
\end{array}
\right.
\end{equation}
The model similar to (\ref{eq_carre_smpl_step}) is used in several rise-and-fall gravimeters \cite{carre1991, alasia1982, murakami1987, zabek2004}. The model is non-linear with parameters $z_a, t_a, g_a, d$ that can be estimated by methods of non-linear regression \cite{bard1973, draper1998}. These estimates give rise to the gravimeter measurement equations, so the properties of the estimates essentially influence characteristics of the instruments.

\section{General properties of linear and non-linear models}
\label{ch_L_NL}
The models like (\ref{eq_3_param}) are \emph{linear} as they represent linear combination of the parameters, in this case $(z_0, V_0, g)$. On the other hand, the models like (\ref{eq_carre_smpl_step}) with parameters $(z_a, t_a, g_a, d)$ are \emph{non-linear}.
    Any model -- linear or not -- agrees with the measured coordinates $(T_i, S_i)$ only up to some errors $\epsilon_i$ caused by both the model incompleteness and data noise, so that
\begin{equation}
\label{eq_model}
S_i = z(T_i) + \epsilon_i.
\end{equation}
In regression analysis $T_i$ are called independent variables, $S_i$ are called observations. The terms do not necessarily reflect the actual way the data are collected. In the IMGC-02 instrument the intervals $S_i$ are pre-determined, while the intervals $T_i$ are measured. The formula (\ref{eq_model}) just assumes that $T_i$ are exactly known, and all the errors (including those arising from this suggestion) are attributed to the values of $S_i$.

Most often the model parameters are estimated by the least-square (LS) solution of the system (\ref{eq_model}) that minimizes the sum of the squared errors: $\sum  \epsilon_i^2 \rightarrow $ min. The estimate for $g$ is the basis of the gravimeter measurement equation, so the properties of the estimates directly transfer to the characteristics of the instrument. We use the following notation for the LS-estimates of $g$:
\begin{eqnarray}
\label{eq_GTS}
\overline g &= G(T_i, S_i), \quad \rm{for\; a\; linear\; model\; (LM)},\\
\label{eq_XTS}
\overline g &= \Xi(T_i, S_i), \quad \rm{for\; a\; non-linear\; model\; (NLM)}.
\end{eqnarray}
The notation highlights the fact that both estimates depend on the measured time-distance coordinates. The similarity, however, ends at this point, as other properties we now consider are significantly different for linear and non-linear models.
\begin{itemize}
\item \emph{Computation.}   The $G(T_i, S_i)$ is a closed-form formula, while $\Xi(T_i, S_i)$ is an iterative process.
\item \emph{Uniquness}. The $G(T_i, S_i)$ produces single value, while the solution for $\Xi(T_i, S_i)$  can be non-unique.
\item \emph{Traceability}. The coordinates $(T_i, S_i)$ are quantities in the base units of time and length referenced to realizations of the primary standards. The solution (\ref{eq_GTS}) for the linear model, being a closed-form one-step formula, provides a direct link between the base quantities and the derived quantity of the acceleration. The measurement errors of the time and length intervals can be explicitly propagated to the uncertainty of the $g$. In contrast, the iterative process (\ref{eq_XTS}) for the non-linear model may not converge or converge to unrealistic solution, depending on many factors, like initial approximations, the method of iterations, the measurement noise, etc. As result, the linear models provide a better-defined link between the primary standards and the measured gravity acceleration.
\item \emph{Linear combination of observations}. LS-estimate for a linear model is always a linear combination of observations:
\begin{equation}
\label{eq_GTS_lin_comb}
\overline g = G(T_i, S_i) = \sum a_i \, S_i.
\end{equation}
For example, for the model (\ref{eq_3_param})
\begin{equation}
\label{eq_GTS_ai}
a_i =
2\left|
\begin{array}{ccc}
N & \sum T_{i} & 1 \\
\sum T_{i} & \sum T_{i} ^{2} &  T_{i} \\
\sum T_{i}^{2} & \sum T_{i}^{3} &  T_{i}^{2}
\end{array}
\right| :\left|
\begin{array}{ccc}
N & \sum T_{i} & \sum T_{i} ^{2} \\
\sum T_{i} & \sum T_{i} ^{2} & \sum T_{i} ^{3} \\
\sum T_{i}^{2} & \sum T_{i}^{3} & \sum T_{i}^{4}
\end{array}
\right|.
\end{equation}
The LS-estimates for non-linear models do not possess this property.
\item \emph{Bias}.
If the errors $\epsilon_i$ are independent, equi-dispersed, and average to zero, then the expectation of the linear estimate equals the expectation of the parameter itself, while the expectation of the non-linear one does not equal the expectation of the parameter:
\begin{equation}
\label{eq_bial}
\mathbb{E}\left (G(T_i, S_i)\right ) = \mathbb{E}(g), \quad
\mathbb{E}\left ( \Xi(T_i, S_i)\right ) \ne \mathbb{E}(g),
\end{equation}
which means that $\Xi(T_i, S_I)$ is always biased.
\item \emph{Separability of influences}. If observations $S_i$ include a component $\Delta S_i$, so that the refined observations $\tilde S_i$ would be
\begin{equation}
\label{eq_S_i_refined}
\tilde S_i = S_i - \Delta S_i,
\end{equation}
then according to the property (\ref{eq_GTS_lin_comb}),
\begin{equation}
\label{eq_S_i_refined}
G(T_i, \tilde S_i + \Delta S_i) =
G(T_i, \tilde S_i ) +
G(T_i, \Delta S_i) = g + \Delta g,
\end{equation}
where
\begin{equation}
\label{eq_Delta_g}
\Delta g = G(T_i, \Delta S_i)
\end{equation}
is the portion of the estimate caused by the influence component $\Delta S_i$. So, for linear models all the influences can be separately analyzed and the corrections obtained with the same computational procedure (\ref{eq_GTS}) used to calculate $g$. This is an important consequence of the superposition principle valid only for linear models. It allows to account for newly recognized influences (e.g. self-attraction or diffraction effects) without invalidating  earlier estimates for $g$.  In contrast, to account for new influences in non-linear case, the influences have to be included in the model and new iteration process (\ref{eq_XTS}) to be run.
\end{itemize}

\section{Non-linear model of the IMGC-02 gravimeter}
\label{sec:NLmodel}
The IMGC-02 absolute gravimeter \cite{bich2008} was built in 2002 by the Istituto Nazionale di Ricerca Metrologica (INRIM) in replacement of the older IMGC instrument \cite{alasia1982} that was in service for over 20 years.  Like several other rise-and-fall gravimeters, the older instrument implemented non-linear model of the trajectory based on (\ref{eq_carre_smpl_step}), with additional terms for the vertical gravity gradient and velocity-proportional accelerations:
\begin{equation}
\label{eq_carre}
{\hat z}(t) = z_a + \frac{g_a}{2}(t - t_a)^2 - \frac{\phi}{6}(t - t_a)^3 + \frac{\gamma}{24}(t - t_a)^4 + d\,{\rm u}(t - t_a),
\end{equation}
where $t_a, z_a$ are coordinates of the apex, $g_a$ is gravity acceleration at the apex, $\phi$ is the rate of the acceleration change with velocity, $\gamma$ is the vertical gravity gradient, $d$ is the residual displacement of the right branch, ${\rm u}(t)$ is the Heaviside step function.
The model of the new IMGC-02 instrument was augmented with four additional parameters to account for the harmonics $\omega_1$ and $\omega_2$ caused by the laser modulation and the resonance frequency of the seismometer:
\begin{eqnarray}
\label{eq_IMGC-02}
{\hat z}(t) & = z_a + \frac{g_a}{2}(t - t_a)^2 - \frac{\phi}{6}(t - t_a)^3 + \frac{\gamma}{24}(t - t_a)^4 + d\,{\rm u}(t - t_a) \nonumber \\
&+ A_1 \sin(\omega_1 t) + B_1 \cos(\omega_1 t) + A_2 \sin(\omega_2 t) + B_2 \cos(\omega_2 t),
\end{eqnarray}
where $A_1, B_1, A_2, B_2$ are sine and cosine amplitudes of the $\omega_1$ and $\omega_2$ corresponding to the known frequencies of 1.18 kHz and 22 Hz \cite{bich2008}. Total, there are ten parameters estimated in every drop of the IMGC-02 gravimeter using the linearization method \cite{dagostino2005}:
\begin{equation}
\label{eq_NL_parms}
g_a \quad t_a \quad z_a \quad \phi \quad \gamma \quad d \quad A_1\quad  B_1 \quad A_2 \quad B_2 
\end{equation}

After the adjustment, the estimated gravity $g_a$ is reduced down the apex at the distance $h_b$:
\begin{equation}
\label{eq_g_b}
g_b = g_ a +  \gamma \, h_b.
\end{equation}
The vertical gradient $\gamma$ used for the reduction is also taken from the adjustment.
The distance $h_b$ known as \emph{best reference height} is derived from the equation of the gradient-perturbed gravity
\begin{equation}
\label{eq_g_h}
g_h = g_a + \gamma \, h
\end{equation}
by taking variances of both sides and minimizing ${\rm var}(g_h)$ over $h$, which yields \cite{carre1991}:
\begin{equation}
\label{eq_h_b}
h_b = {\rm cov}(g_a, \gamma)/{\rm var}^2(\gamma).
\end{equation}
The variances and covariances are found in every drop as part of the linearization algorithm.
It was experimentally confirmed \cite{bich2008} that the variance of the value $g_b$ (\ref{eq_g_b}) is several times less than the variance or $g_a$. For this reason the value $g_b$ is reported as measurement result at the height $h_b$ down the apex. 
The best measurement height (\ref{eq_h_b}) is only used in nonliniear models of rise-and-fall gravimeters and should not be confused with the \emph{effective measurement height} known for the linear models \cite{niebauer1989, timmen2003}. Even though we found that both heights can be pretty close together, there are concerns regarding the implementation of the best measurement heigh.
The gradient coming from the non-linear fit can take values outside any realistic range\footnote{like in the table \ref{tab_NL}}, so the calculations done in (\ref{eq_g_b}) can hardly represent downward reduction of the gravity at the distance $h_b$. It is not completely clear how the combination of the larger uncertainty values of $g_a$ and $\gamma$ in (\ref{eq_g_b}) leads to the value $g_b$ with much smaller uncertainty. 
The equation (\ref{eq_h_b}) provides only a partial explanation, because it assumes that the gradient is the only disturbance of the trajectory (\ref{eq_g_h}). But in fact the gradient is only one of the nine other parameters (\ref{eq_NL_parms}) and represents neither biggest, nor most uncertain disturbance.

Despite some open questions, the non-linear model (\ref{eq_IMGC-02}) produces stable results for most drops. However, for the drops with high data noise the regression algorithm may fail or converge to unrealistic values. The non-linear model also complicates accounting for disturbances not included in the model, which limits operational capabilities and impedes furter improvements of the instrument.
\section{Linear Model}
\label{sec:Lmodel}
We implement the linear model in three steps. First, we recover the right branch of the trajectory by reflecting it down with respect to the apex, like in (\ref{eq_refl_trans}), and obtain the vector of coordinates $\mathbf{S}=\{S_1, ..., S_N\}^T$ that matches the time vector $\mathbf{T}= \{T_1, ..., T_N\}^T$.

Second, we fit the simple parabola (\ref{eq_3_param}) to the recovered trajectory using the standard LS formula
\begin{equation}
\label{eq_LSS}
\mathbf{\xi} = \mathbf{A}^+ \mathbf{S},
\end{equation}
where
\begin{equation}
\mathbf{\xi} =
\left(
\begin{array}{c}
S_0 \\
V_0 \\
g
\end{array}
\right)
, \quad
\mathbf{A}^+=(\mathbf{A}^T\mathbf{A})^{-1}\mathbf{A}^T
, \quad
\mathbf{A} =
\left(
\begin{array}{ccc}
1       & T_1    & T_1^2/2 \\
1       & T_2    & T_2^2/2 \\
\vdots  & \vdots & \vdots  \\
1       & T_{N}  & T_{N}^2/2
\end{array}
\right)
\label{eq_LSS_expanded}
\end{equation}
Note that the solution for $g$ given by the above formula (\ref{eq_LSS}) is the same as by formula (\ref{eq_GTS_lin_comb}), as both formulas are different methods (the Cramer's rule and the Moore-Penrose pseudoinverse) of solving the same system of normal equations, and so the factors $a_i$ of (\ref{eq_GTS_ai}) are found in the third row of the pseudoinverse matrix $\mathbf{A}^+$.

On the third step we apply necessary corrections to the result. To calculate the corrections we refine the recovered trajectory (\ref{eq_refl_trans}) by subtracting the known disturbances:
\begin{equation}
\label{eq_Si_refined}
\tilde S_i = S_i
-  \Delta S_i^d
-  \Delta S_i^{\gamma}
-  \Delta S_i^{\phi}
-  \Delta S_i^{\omega 1}
-  \Delta S_i^{\omega 2}
,
\end{equation}
where upper indexes stand for the corresponding disturbances. In many cases it's more convenient to analyze disturbances in terms of acceleration rather than coordinate, which can be done using the gravimeter weighting functions. In the following sections we discuss the weighting functions of the IMGC-02 gravimeter and apply them for the analysis of the disturbances found in (\ref{eq_Si_refined}).
\subsection{Weighting functions of the IMGC-02 gravimeter}
\label{wf}
For the rise-and-fall gravimeters it is convenient to relate the time origin to the apex of the trajectory, so that the measurement interval would change from $-T/2$ to $T/2$ rather than from 0 to $T$. The translation is
\begin{equation}
\label{eq_Ti_centering}
\hat{T}_i=T_i-t_a.
\end{equation}
When calculati  ng $g$ using the formulas (\ref{eq_LSS}, \ref{eq_LSS_expanded}) or (\ref{eq_GTS_lin_comb}, \ref{eq_GTS_ai}) the time shift can be ignored, because the estimates for the quadratic coefficient of the simple parabola  (\ref{eq_3_param}) are not sensitive to the time shift, so that
\begin{equation}
\label{eq_GTS_time_shift}
G(\hat{T}_i  \, , \, S_i) = G(T_i  \, , \, S_i).
\end{equation}
However, for the analysis of the disturbances the shift is important, as the disturbances rarely follow the simple parabolic shape. We will use the notation $T_i \uparrow$ or $T_i \downarrow$ for the values associated with only upward or downward branches of the trajectory.

\begin{figure}
\begin{center}
\begin{tabular}{@{}c@{}}
\includegraphics[height=60mm]{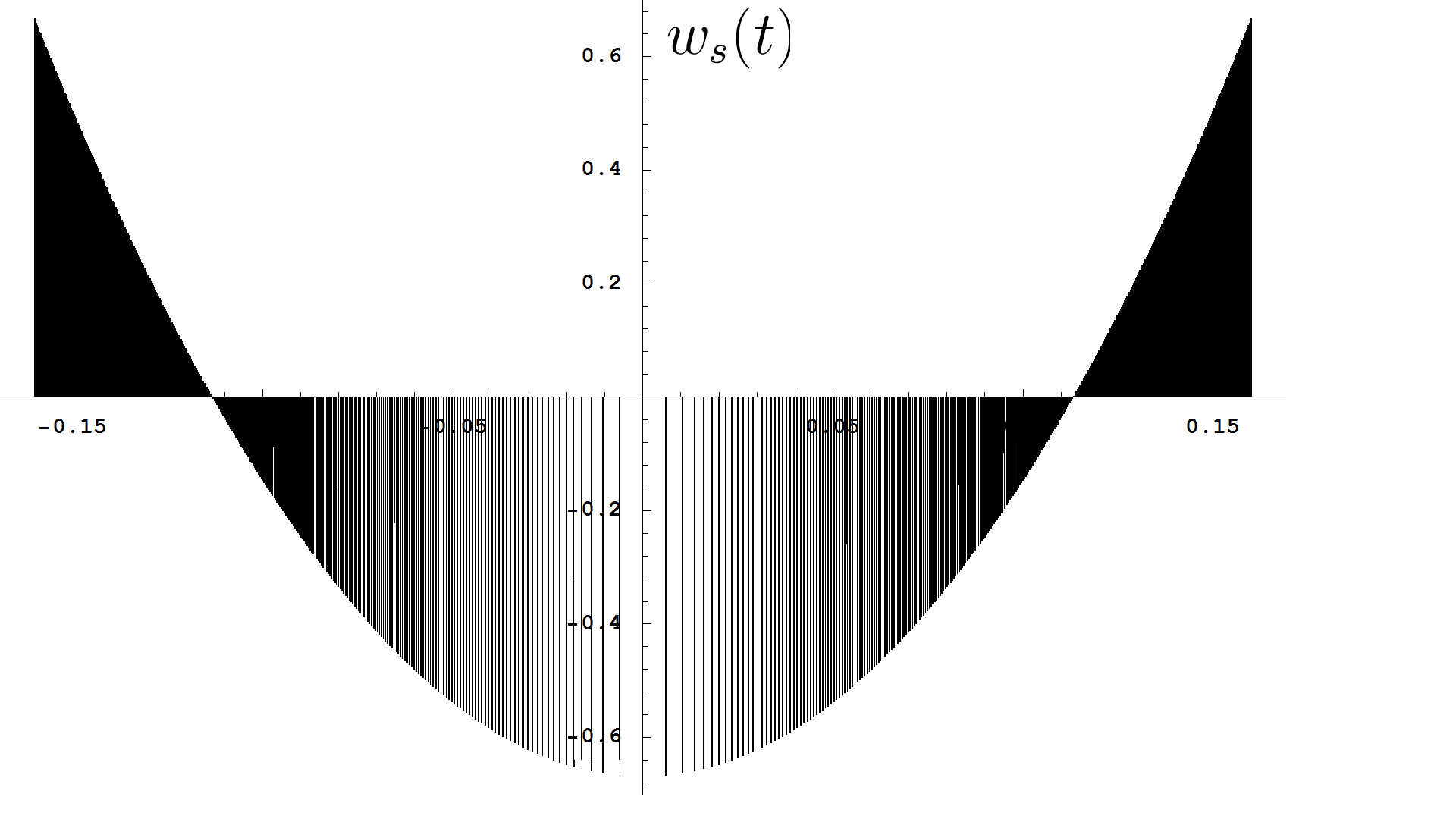} \\
\includegraphics[height=60mm]{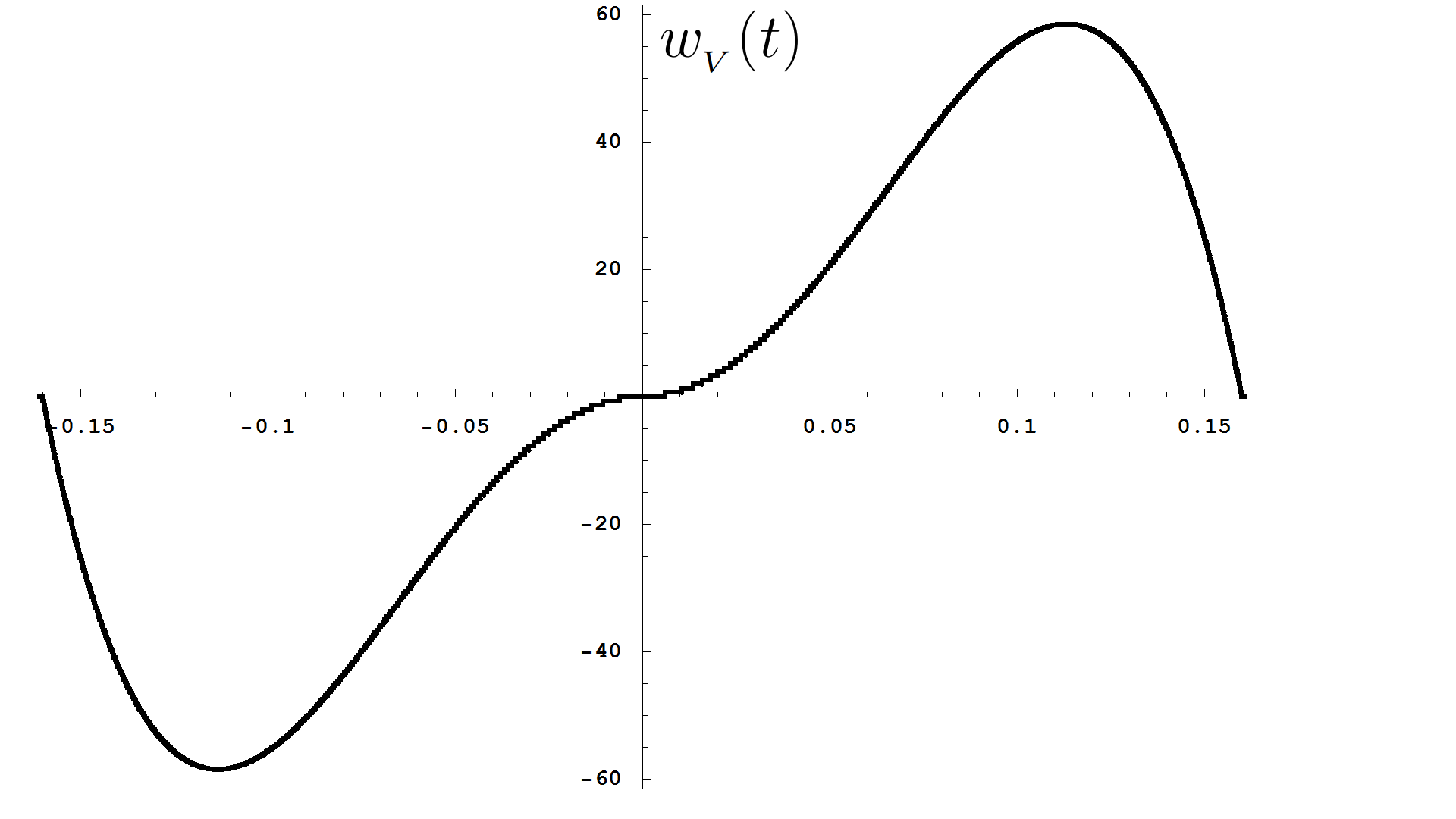} \\
\includegraphics[height=60mm]{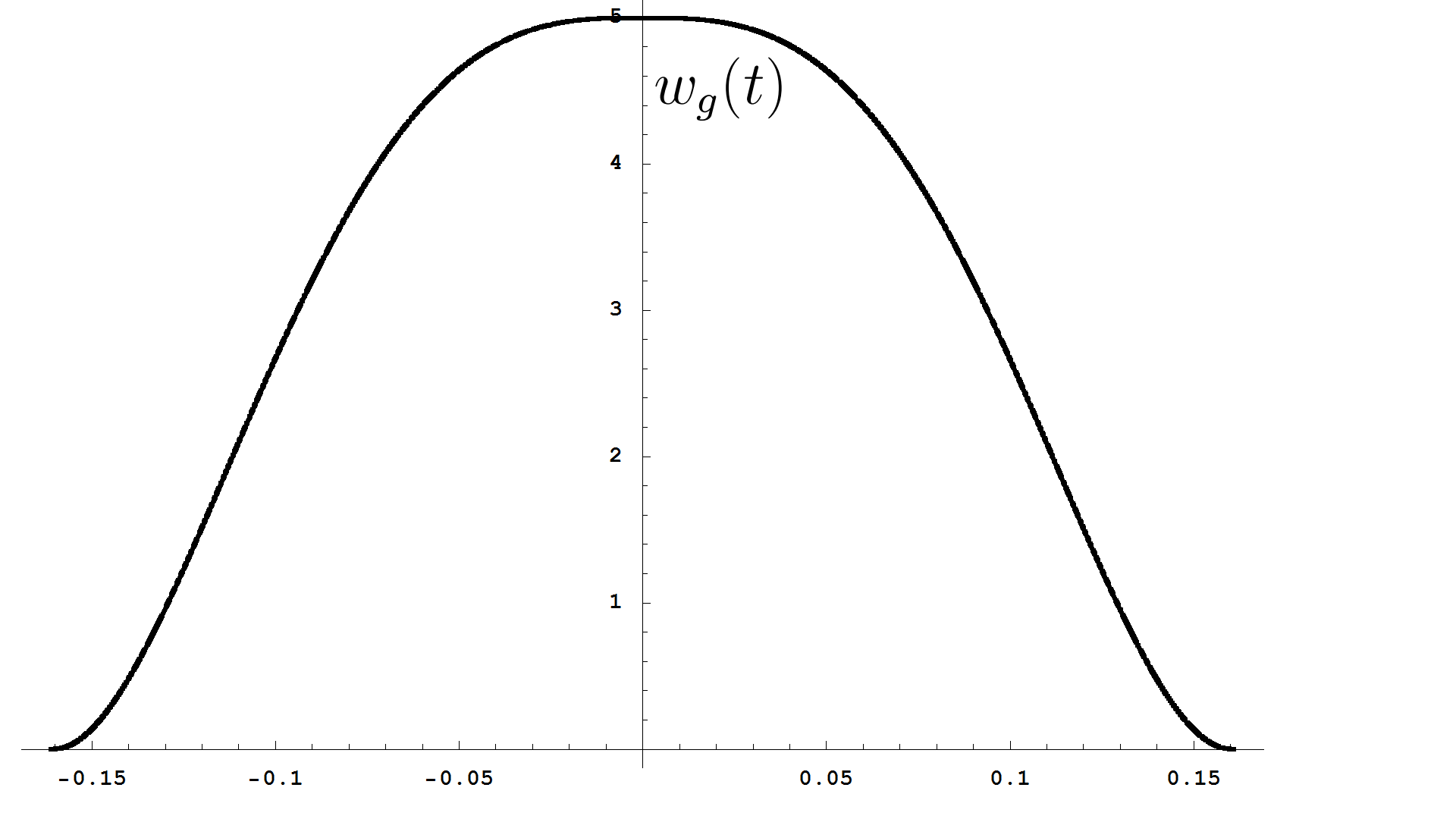} \\
\end{tabular}
\end{center}
\caption{Weighting functions of IMGC-02 gravimeter by coordinate, velocity, acceleration in case of joint processing of both parts of the trajectory.}
\label{fig_WFs_sym}
\end{figure}

The formula (\ref{eq_GTS_lin_comb}) presents the LS-estimate as linear combination of observations. Every observation $S_i$ can be viewed as sampling of continuously changing coordinate $S(t)$ taken at the moment $T_i$. Using the Dirac delta-function $\delta (t)$ this can be written as
\begin{equation}
\label{eq_Si_via_ddelta}
S_i = \int_{-T/2}^{T/2} S(t) \,  \delta(t-T_i) \D t .
\end{equation}
Substituting this $S_i$ into (\ref{eq_GTS_lin_comb}) we get
\begin{equation}
\label{eq_Si_via_ws}
\overline g =  \int_{-T/2}^{T/2} S(t) \sum a_i \, \delta(t-T_i) \D t  =  \int_{-T/2}^{T/2} S(t) \, w_s(t)\D t.
\end{equation}
We thus expressed the measured gravity as weighted average of the test mass coordinates, where the weights are defined by the function\footnote{The formula (\ref{eq_ws}) in conjunction with the relationships (\ref{eq_ws_wv_wg}) represents an alternative treatment of the weighting function approach \cite{nagornyi1995}, leading to the same results.}
\begin{equation}
\label{eq_ws}
w_s(t) =  \sum a_i \, \delta(t-T_i).
\end{equation}
This function (fig.~\ref{fig_WFs_sym}) is a sequence of $N$ delta-impulses applied at the moments $T_i$ and scaled by the factors $a_i$ (\ref{eq_GTS_ai}).
The change of the measured gravity caused by the disturbance $\Delta S(t)$ is
\begin{equation}
\label{eq_dg_via_ws}
\Delta \overline g = \int_{-T/2}^{T/2} \Delta S(t) \, w_s(t)\D t.
\end{equation}
The change can also be expressed in terms of velocity $\Delta V(t)$ or acceleration $\Delta g(t)$ as
\begin{equation}
\label{eq_dg_via_wv}
\Delta \overline g = \int_{-T/2}^{T/2} \Delta V(t) \, w_v(t)\D t,
\end{equation}
\begin{equation}
\label{eq_dg_via_wg}
\Delta \overline g = \int_{-T/2}^{T/2} \Delta g(t) \, w_g(t)\D t.
\end{equation}
The weighting functions  $w_g(t)$, $w_{{}_V}(t)$, and $w_s(t)$ are related as \cite{nagornyi2013}
\begin{equation}
\label{eq_ws_wv_wg}
w_s(t) = -\frac{\D}{\D t}w_{{}_V}(t) =
\frac{\D^2}{\D t^2} w_g(t).
\end{equation}
The following properties are always hold true:
\begin{equation}
\label{eq_wz_wv_wg_square}
\int_{-T/2}^{T/2} \!\!\!  w_g(t) \D t = \!\! 1, \quad
\int_{-T/2}^{T/2} \!\!\!  w_{_V}(t) \D t = 0, \quad
\int_{-T/2}^{T/2} \!\!\!  w_s(t) \D t = 0.
\end{equation}
The function $w_s(t)$ defines weights implicitly applied by the gravimeter LS-procedure to the measured coordinates to obtain the acceleration\footnote{These weights should not be confused with those explicitly applied to observations in weighted least-squares (WLS) estimates.}. The function $w_g(t)$ shows weights that gravimeter applies directly to the test mass acceleration. If disturbance of the acceleration is expanded like
\begin{equation}
\label{eq_g_t_exp}
\Delta g(t) = \sum b_n t^n,
\end{equation}
then the measured gravity, according to (\ref{eq_dg_via_wg}), changes like
\begin{equation}
\label{eq_dg_via_Cn}
\Delta \overline {g} = \sum b_n C_n,
\end{equation}
where $C_n$ are found as \cite{nagornyi1995}
\begin{equation}
\label{eq_Cn}
C_n = \int_{-T/2}^{T/2} t^n w_g(t) \D t = \frac{G(\hat{T}_i, \hat{T}_i^{n+2})}{(n+1)(n+2)}.
\end{equation}
The asymptotic ($N \rightarrow \infty$) weighting function $w_g(t)$ of the IMGC-02 gravimeter can be found by the limiting transition in the sums (\ref{eq_GTS_ai}) leading to
\begin{equation}
\label{eq_wg_sym_ESD}
w_g(t) = \frac{76.8}{T^6} t^4|t| - \frac{32}{T^4} t^2|t| + \frac{1.6}{T}.
\end{equation}
The details of this transition are given in the Appendix. The limiting values for the coefficients $C_n$ are found by the direct substitution of (\ref{eq_wg_sym_ESD}) to the integral in (\ref{eq_Cn}) leading to
\begin{equation}
\label{eq_Cn_sym_ESD}
C_n = \left\{
\begin{array}{ll}
24\,(T/2)^n (n^3 + 11n^2 + 34n + 24)^{-1}    &,   \,  n \rm{\; is\; even},
\\
0    &,   \, n \rm{\; is\; odd}.
\end{array}
\right.
\end{equation}
Zeroes for odd $n$ are due to the central symmetry of the $w_g(t)$. For the finite number of levels the symmetry may not be perfect, in which case more generic formula (\ref{eq_Cn}) has to be used.

\begin{figure}
\begin{center}
\begin{tabular}{@{}ccc@{}}
$w_z$ & \includegraphics[height=40mm]{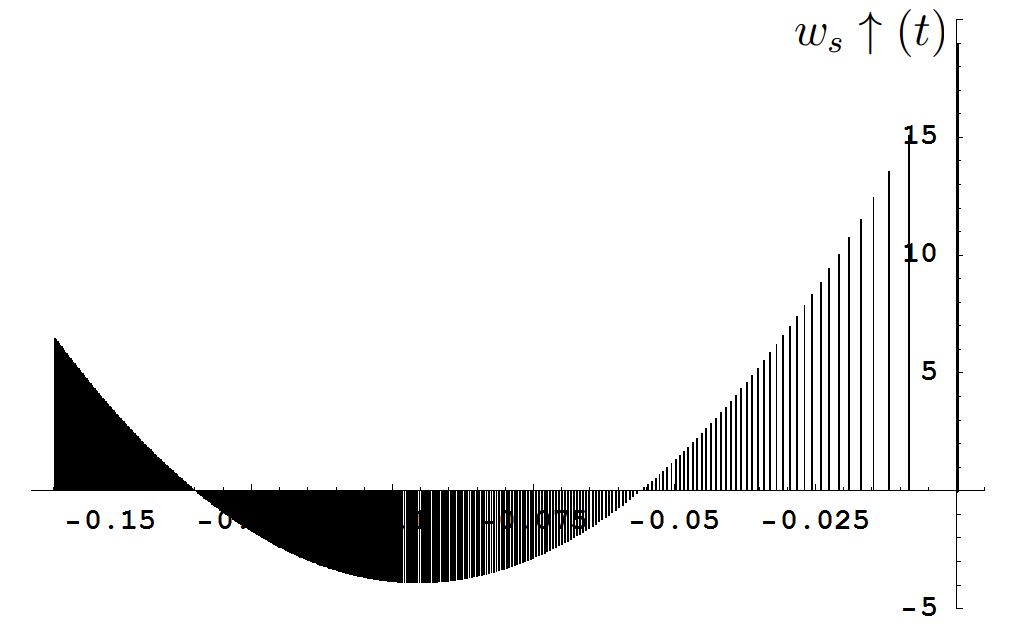} & \includegraphics[height=40mm]{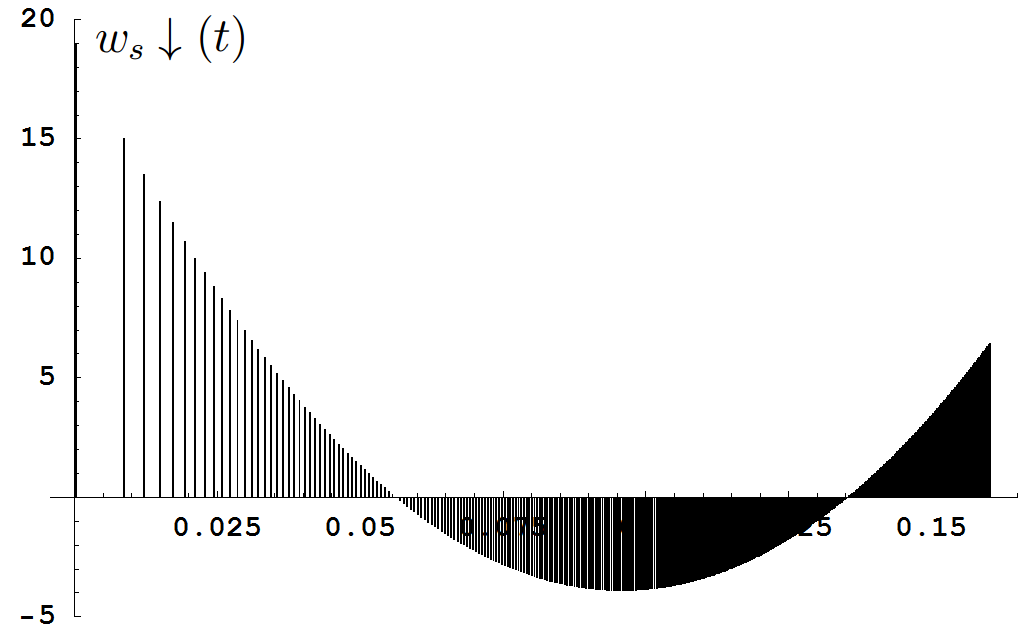} \\
$w_{{}_V}$ & \includegraphics[height=40mm]{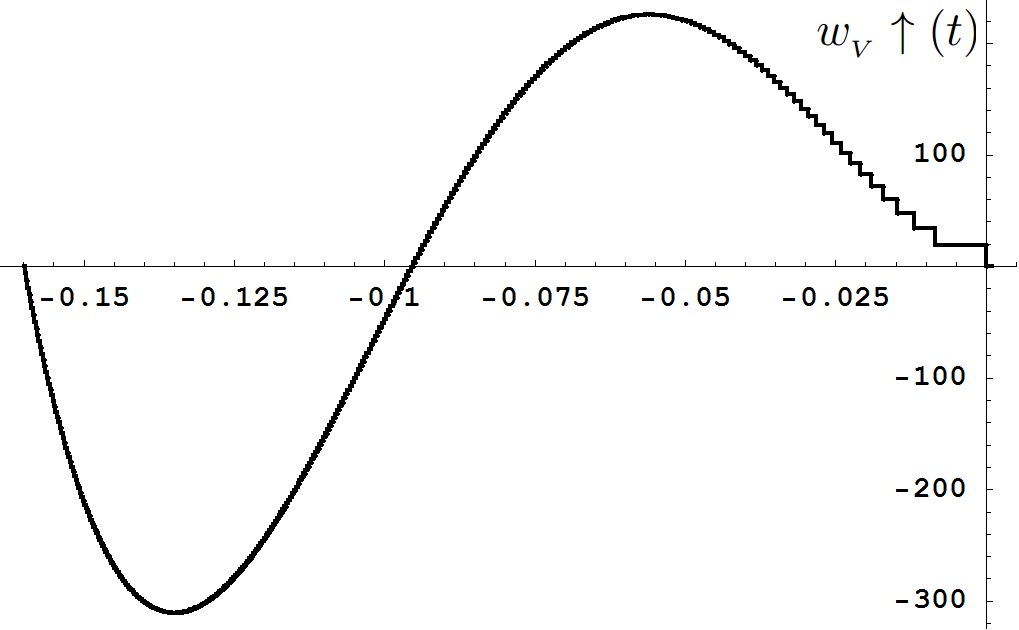} & \includegraphics[height=40mm]{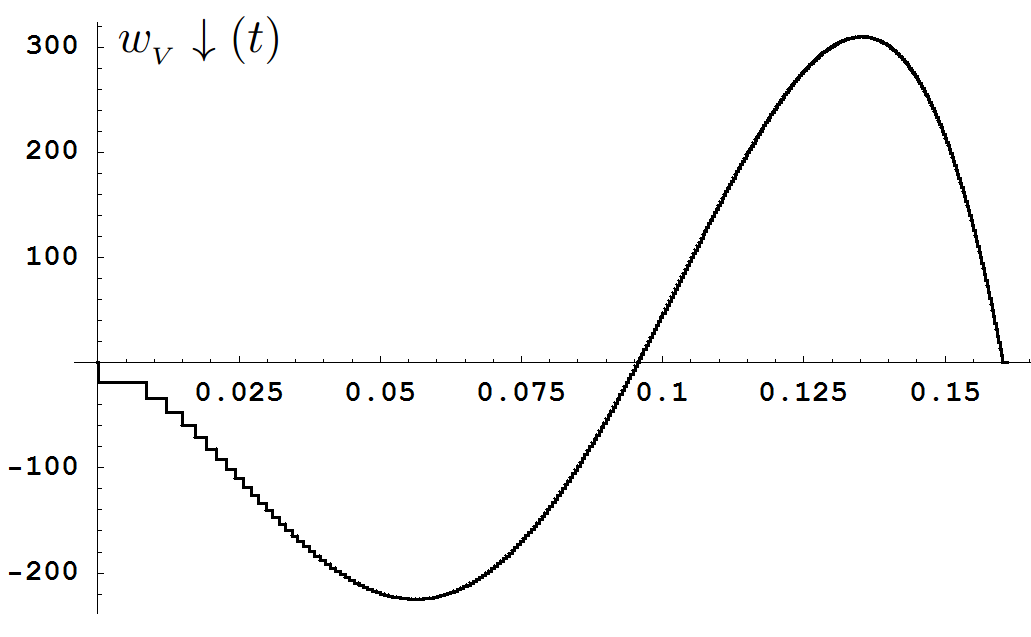} \\
$w_g$ & \includegraphics[height=40mm]{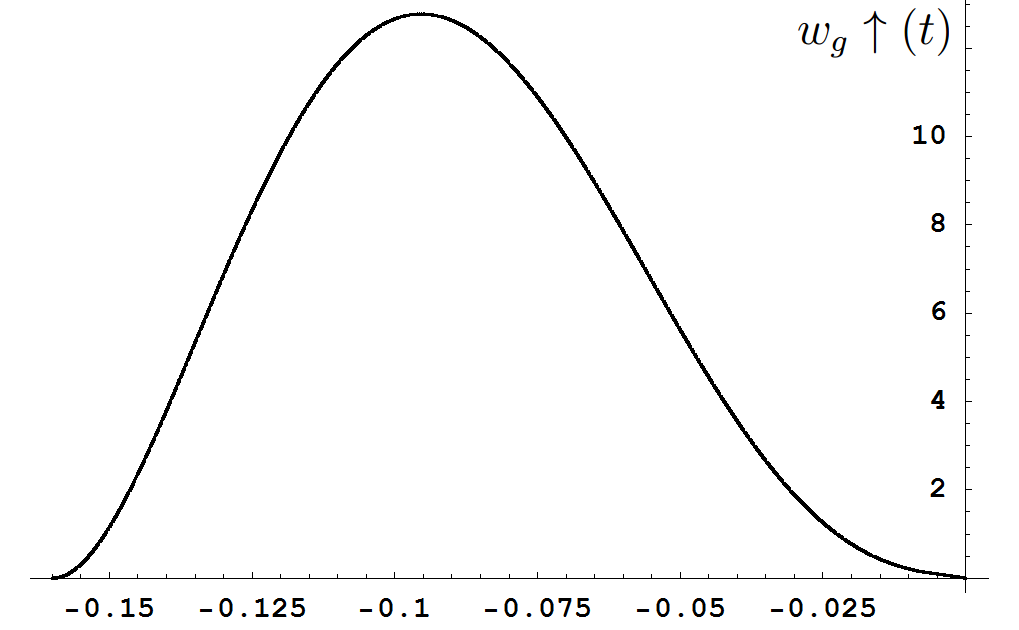} & \includegraphics[height=40mm]{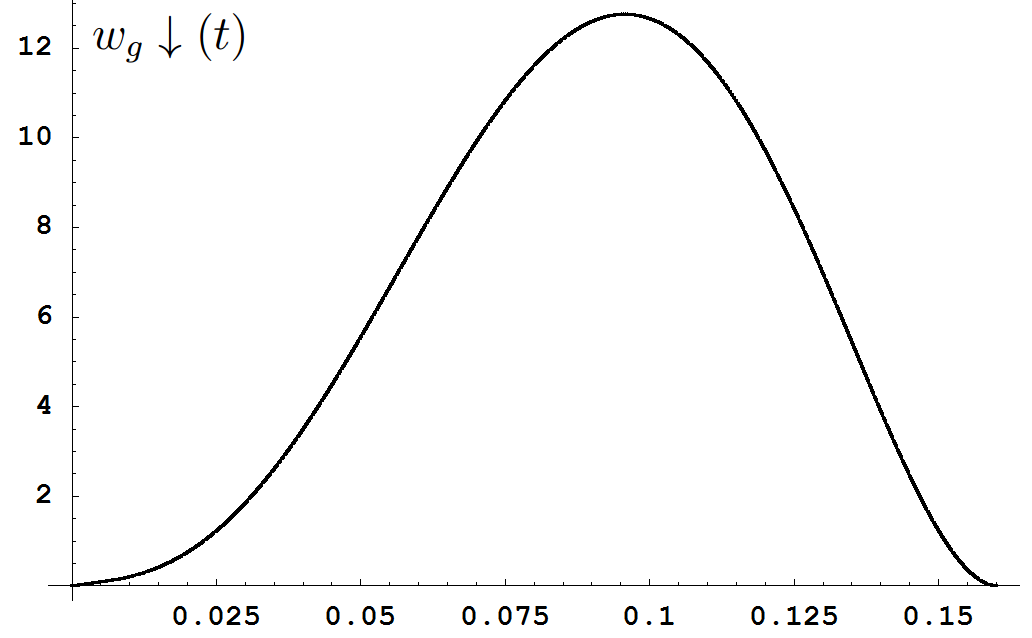}  \\
\end{tabular}
\end{center}
\caption{Weighting functions of IMGC-02 gravimeter by coordinate, velocity, acceleration in case of separate processing of the upward and the downward parts of the trajectory}
\label{fig_WFs_half}
\end{figure}

If the parabola is fitted separately to the left and right branches of the trajectory, the weighting functions of the gravimeter will change (fig.\ref{fig_WFs_half}).
For the downward branch
\begin{equation}
\label{eq_wg_down}
w_g(t) \downarrow = \frac{60}{(T/2)^6} t^5   -\frac{120}{(T/2)^5} t^4     + \frac{60}{(T/2)^4} t^3,
\end{equation}
which corresponds to the weighting function of the direct free-fall gravimeter with the levels equally spaced in distance (ESD) and the measurement interval of $T/2$ \cite{nagornyi1995}. The moments $C_n$ of this weighting function are
\begin{equation}
\label{eq_Cn_half}
C_n  \downarrow = 120\,(T/2)^n (n^3 + 15 n^2 + 74 n + 120)^{-1} .
\end{equation}

\subsection{The apex step}
\label{sec_apex_step}
\begin{figure}
\begin{center}
\begin{tabular}{@{}c@{}}
\includegraphics[height=40mm]{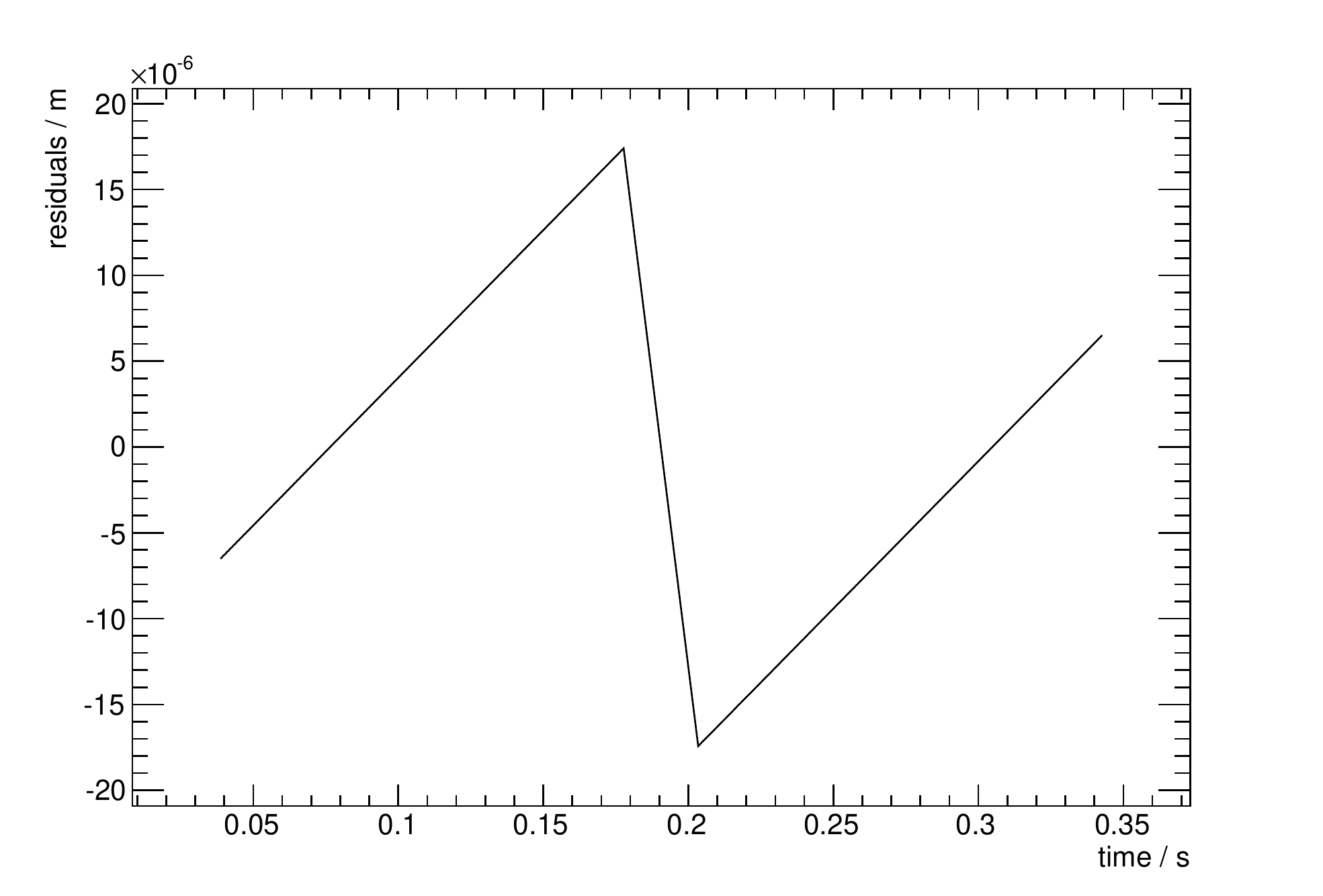}   \\
\includegraphics[height=40mm]{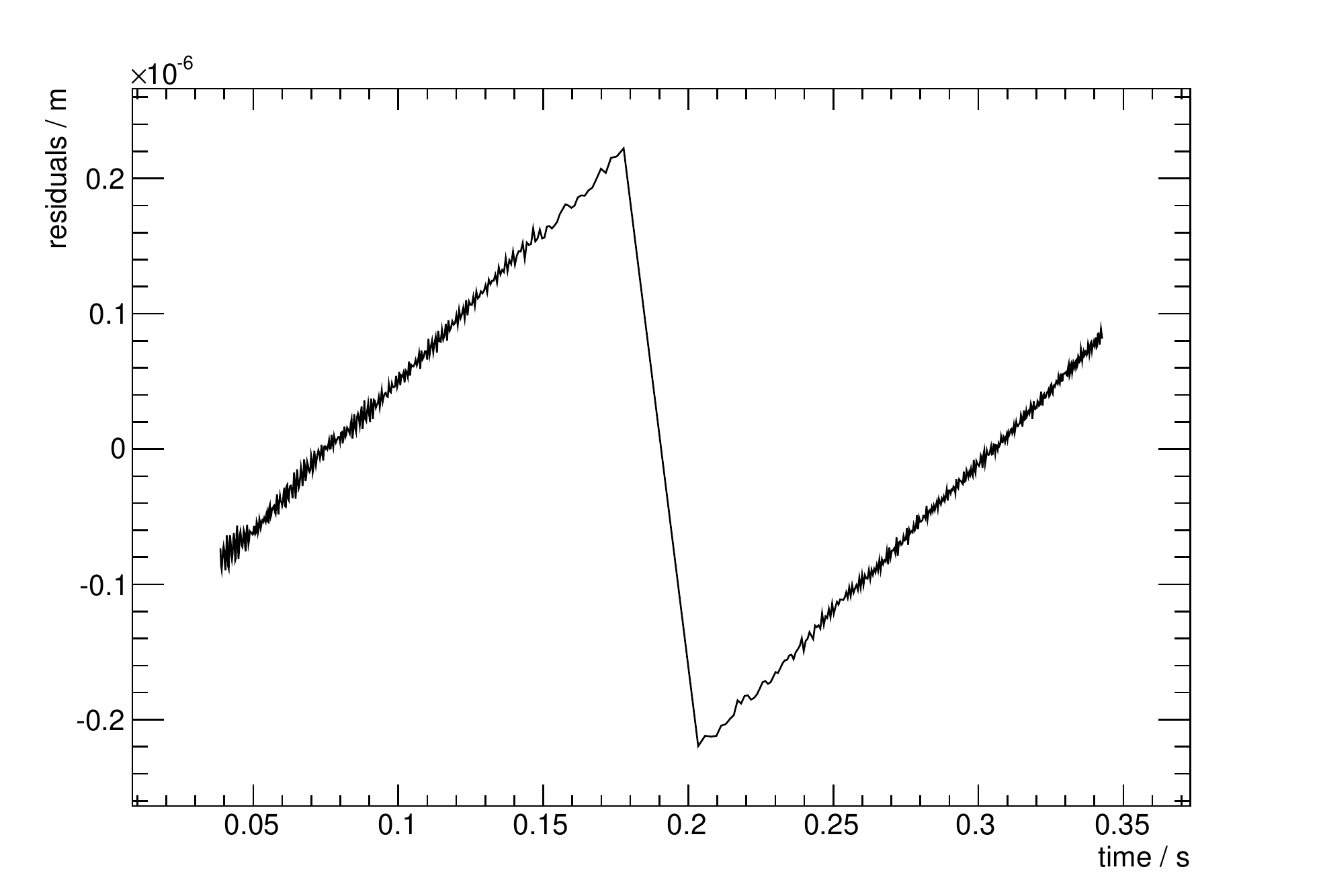} \\
\includegraphics[height=40mm]{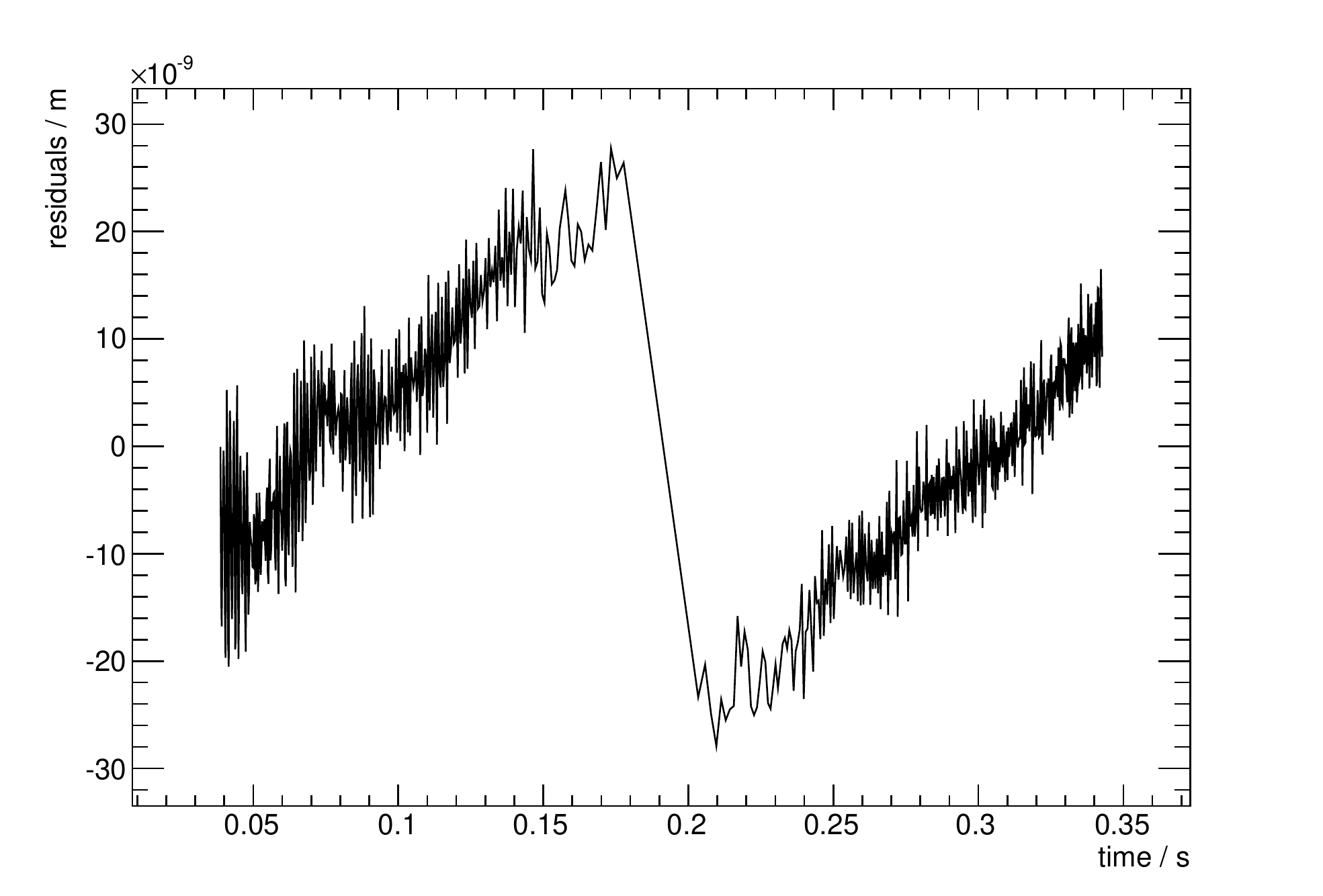} \\
\includegraphics[height=40mm]{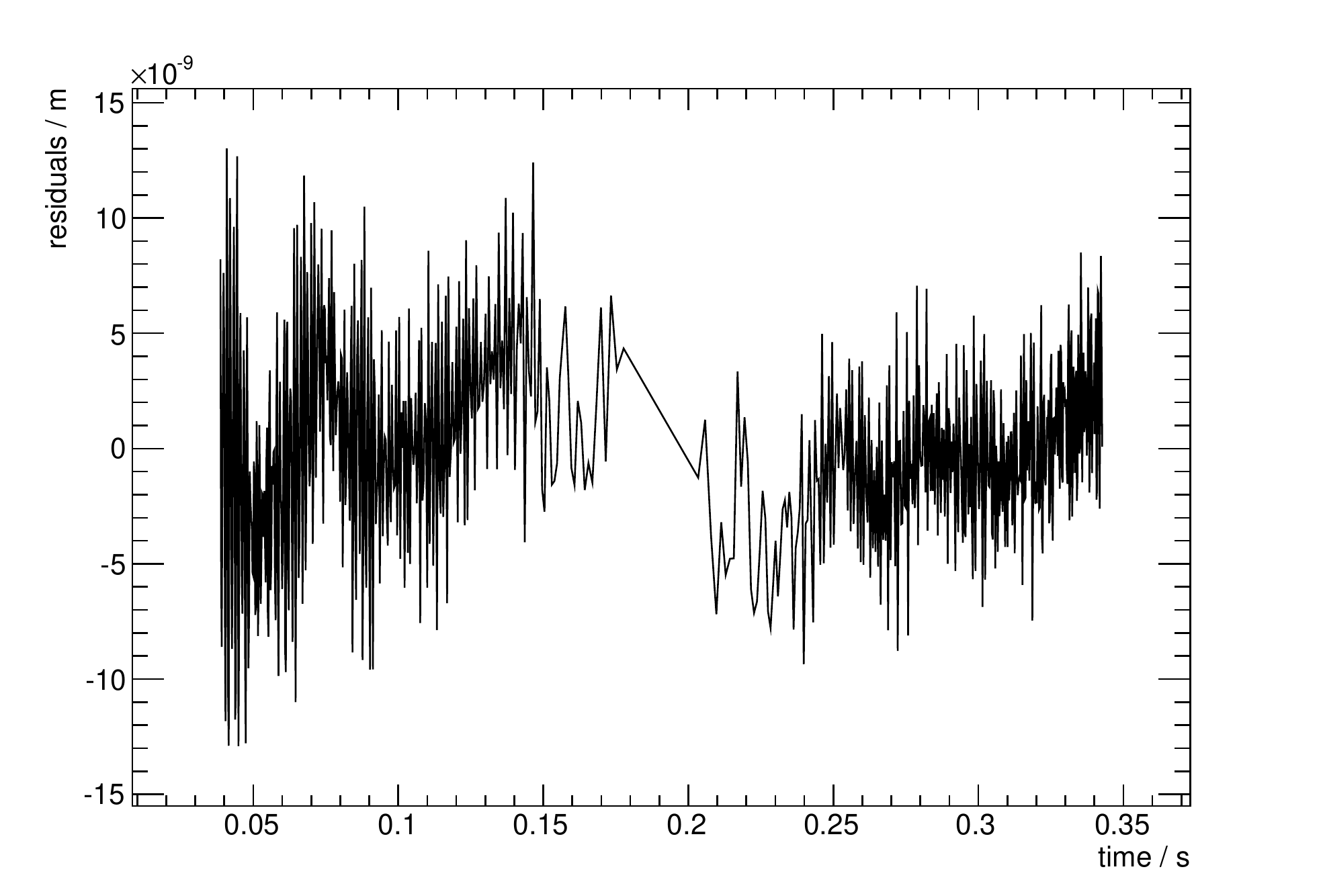} \\
\includegraphics[height=40mm]{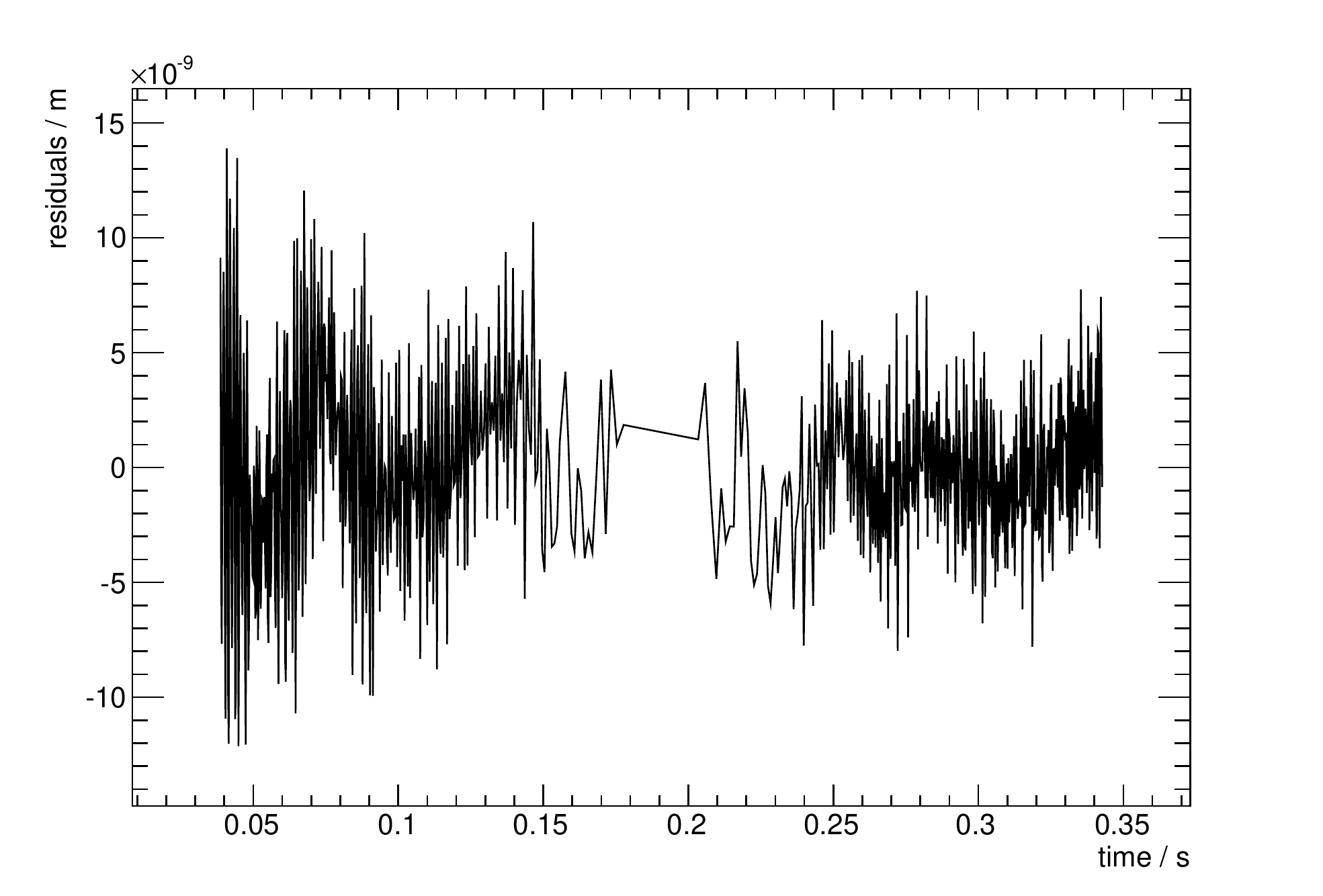}
\end{tabular}
\end{center}
\caption{Iterative reduction of the apex step}
\label{fig_apex_iter}
\end{figure}

We restore the right branch of the trajectory using the formula (\ref{eq_refl_trans}). The approximate coordinates $(t_a, z_a)$ can be found in several ways, such as
\begin{itemize}
\item fitting separate parabolas to the left and right branches;
\item finding the minimum of the velocity by numerical differentiation and smoothing the trajectory;
\item use of non-linear models.
\end{itemize}
As the apex is not exactly known, the residuals of fitting the parabola to the restored trajectory reveal the step discussed in the section \ref{sec_apex_singularity}. We eliminate the step by vertical adjustment of the right branch, and then do the fit again. The process is repeated until the step is undistinguished in the residuals (fig.~\ref{fig_apex_iter}).
Because of the data noise, the final position of the right branch will retain some uncertainty, but not exceeding one-half of the range or the residuals:
\begin{equation}
\label{eq_res_step}
d < \frac12 | \max \epsilon_i - \min \epsilon_i |.
\end{equation}
For noisy sites the value of $d$ may reach  $10^{-8}$ m. To assess how this step affects the measured gravity we consider the disturbance in the form of the unit step applied at the apex:
\begin{equation}
\label{eq_dS1}
S_i^{01} = \left\{
\begin{array}{ll}
0 &  , \,  T_i< t_a, \\
1  &  , \,  T_i > t_a.
\end{array}
\right.
\end{equation}
The actual apex step differs from (\ref{eq_dS1}) only by the factor $d$, so the reaction of the gravimeter on the step can be found as \cite{nagornyi1993}
\begin{equation}
\label{eq_Delta_gd}
\Delta g_d = d \cdot G(T_i, S_i^{01}) = d \; \frac{\D w_g(t_a)}{\D t}.
\end{equation}
The factor $\D w_g(t_a) / \D t$ is the derivative of the weighting function $w_g(t)$ at the apex. The derivative turns to zero in case of the perfect symmetry of the $\{T_i\}$-set with respect to the apex $t_a$. In real drops of the IMGC-02 gravimeter the apex derivative is distributed about uniformly within $\pm$ 0.25 s$^{-2}$, so the remaining portion of the step leads to an uncertainty of less than 0.25~\muG\footnote{1 \muG = $10^{-8}$ ms$^{-2}$} in one drop.

\subsection{Vertical gravity gradient}

Vertical gravity gradient $\gamma$ changes the test mass acceleration proportionally to the mass separation from the apex:
\begin{equation}
\label{eq_gt_grad}
\Delta g_{\gamma}(t) = \gamma \, z(t) = \gamma \, g\, t^2/2.
\end{equation}
According to (\ref{eq_dg_via_Cn}), this disturbance changes the measured gravity like
\begin{equation}
\label{eq_dg_grad}
\Delta \overline g_{\gamma} = 
\gamma \, g \, C_2/2.
\end{equation}
Substituting $C_2 = T^2/24$ from (\ref{eq_Cn_sym_ESD}) we get
\begin{equation}
\label{eq_dg_grad_heff}
\Delta \overline g_{\gamma} = \gamma \, g \, T^2/48 = \gamma \, H/6 = \gamma \, h_{\rm eff}.
\end{equation}
The point below the apex at the distance $h_{\rm eff}$ is called the \emph{effective measurement height} of the gravimeter. In assumption of the constant vertical gradient, the change of gravity from that point to the apex is exactly the same as the correction (\ref{eq_dg_grad}), so without the correction the measured gravity corresponds to the effective measurement height. The value of $H/6$ in (\ref{eq_dg_grad}) is obtained for infinite number of levels equally spaced in distance. In general case the formula (\ref{eq_Cn}) rather than its approximation (\ref{eq_Cn_sym_ESD}) should be used for the coefficient $C_2$ leading to
\begin{equation}
\label{eq_heff}
h_{\rm eff} = \frac{H}{3\,T^2} G(\hat{T}_i, \hat{T}_i^4).
\end{equation} 
\subsection{Velocity-proportional components}
The velocity-proportional components create the following disturbance of the test mass acceleration
\begin{equation}
\label{eq_dgt_phi}
\Delta g_{\phi}(t) = \phi \; V(t) =  \phi \, g \, t,
\end{equation}
where  $\phi$ is the rate of the acceleration change with the velocity. 
There are several effect that contribute to the disturbance (\ref{eq_dgt_phi}), which we present as
\begin{equation}
\label{eq_phi}
\phi = \phi_{_{RG}} + \phi_{_{FSL}}+ \phi_{_{OTH}},
\end{equation}
where $\phi_{_{RG}}$ is the contribution of the residual gas of the vacuum chamber, $\phi_{_{FSL}}$ is the contribution of the finite speed of light, $\phi_{_{OTH}}$ is the contribution of multiple other effects of lower magnitude, such as electrical and magnetic fields, electronics phase delays, etc. \cite{niebauer1995}. 
The residual gas disturbance is always contra-directed to the velocity, is difficult to estimate theoretically, but completely goes away on high vacuum. In contrast, the speed-of-light disturbance is always the same for an instrument, may have different directions depending on the position of the interferometer\footnote{In the IMGC-02 instrument the speed-of-light disturbance is also contra-directed to the velocity, due to the upper position of the interferometer}, and is well defined analytically \cite{nagornyi2011}. For the IMGC-02 instrument
\begin{equation}
\label{eq_phi_fsl}
\phi_{_{FSL}} = -\,3\,g/c,
\end{equation}
where $c$ is speed of light{\footnote{Some researchers believe that the coefficient in the formula (\ref{eq_phi_fsl}) has to be 2 rather than 3. Interested readers can follow the discussion of the subject in publications \cite{rothleitner2011}, \cite{nagornyi2011a}, \cite{rothleitner2011a}, \cite{nagornyi2012}}. The total $\phi$ (\ref{eq_phi}) can be estimated using separate processing of the upward and downward branches of the trajectory. When travelling up and then down, the acceleration of the test mass changes like
\begin{equation}
\label{eq_gt_up}
g \uparrow (t) = g + g_{\phi}(t) + g_{\gamma}(t),
\end{equation}
\begin{equation}
\label{eq_gt_down}
g \downarrow (t) = g - g_{\phi}(t) + g_{\gamma}(t).
\end{equation}
The measured accelerations will be
\begin{equation}
\label{eq_g_up}
\overline g \uparrow  = g - g \, \phi \, C_1 \uparrow  + \overline g_{\gamma},
\end{equation}
\begin{equation}
\label{eq_g_down}
\overline g \downarrow  = g + g \, \phi \, C_1 \downarrow + \overline g_{\gamma},
\end{equation}
where $C_1\downarrow=C_1\uparrow$ are both found with formula (\ref{eq_Cn_half}). The difference $\overline g \uparrow$ and $\overline g \downarrow$ cancels the constant acceleration $g$ and the gradient disturbance $g_{\gamma}(t)$, while doubling the effect of the  $g_{\phi}(t)$. We get
\begin{equation}
\label{eq_phi_calc}
\phi = \frac{\overline g \uparrow - \overline g \downarrow} {2 \, g \, C_1} = 
\frac74 \frac{\overline g \uparrow - \overline g \downarrow} { g \, T } = \frac{7\,T}{32\,H} ( \overline g \uparrow - \overline g \downarrow ) .
\end{equation}
One of the advantages of rise-and-fall gravimeters is their insensitivity to the velocity-proportional disturbances due to the symmetry of the trajectory. The symmetry, however, is never perfect. According to (\ref{eq_dg_via_wg}), the disturbance \label{eq_dgt_phi} causes the following change in the measured gravity
\begin{equation}
\label{eq_dg_phi}
\Delta \overline g_{\phi} = \int_{-T/2}^{T/2} \Delta g_{\phi}(t) \, w_g(t)\D t = \phi \, g  \int_{-T/2}^{T/2} t \, w_g(t)\D t = \phi \, g \, C_1, 
\end{equation}
where $C_1$ is the first central moment of the weighting function $w_g(t)$ found over the entire measurement interval (\ref{eq_Cn}):
\begin{equation}
\label{eq_C1_central}
C_1 = \int_{-T/2}^{T/2} t w_g(t) \D t =  G(\hat{T}_i \; , \;  \hat{T}_i^3 ) / 6.
\end{equation}
In the measurements performed by the IMGC-02 instrument the value of $C_1$ never exceeded $10^{-14}$~s, while $\phi$ stayed below $10^{-6}$ s$^{-1}$, making the influence of the velocity-proportional components (\ref{eq_dg_phi}) really negligible.
\subsection{Harmonic disturbances}

When the gravity value is found by the linear model (\ref{eq_3_param}), the error caused by the harmonic disturbance of the coordinate falls within the range \cite{svitlov2012}
\begin{equation}
\label{eq_dg_harmonic}
\Delta g_{\omega} = \pm \, \omega^2 \, \alpha \, |A(\omega)|,
\end{equation}
where $\alpha, \omega$ are the amplitude and the angular frequency of the disturbance, $A(\omega)$ is the gravimeter amplitude frequency response. The error assumes different values from the range (\ref{eq_dg_harmonic}) depending on the initial phase of the disturbance. For the IMGC-02 gravimeter, the frequency response can be found as Fourier transform of its weighting function (\ref{eq_wg_sym_ESD}) leading to \cite{svitlov2012}
\begin{eqnarray}
A(\omega) = \frac{24}{(\w)^3}
\Bigg(
&- \sin(\w)
-\frac{1+5\cos(\w)}{(\w)}   \nonumber \\
&+ 12 \frac{\sin(\w)}{(\w)^2}
+ 12\frac{\cos(\w)-1}{(\w)^3}
\Bigg) \label{eq_Aw}
.
\end{eqnarray}
The disturbance caused by the laser wavelength modulation has $f=\omega/(2\pi)=1.17$~kHz, $\alpha \approx 3$~nm \cite{dagostino2005}, with random initial phase. According to (\ref{eq_dg_harmonic}), this introduces random error of about $\pm$ 0.3~\muG which is really negligible. The disturbance caused by the seismometer resonance has $f \approx$~22~Hz, $a \approx$ 3~nm \cite{dagostino2005}, introducing a pretty significant error of up to 10~\muG which can be systematic if the initial phase is correlated between drops.

The analysis of the least-squares residuals shows that this disturbance may not be prominent in all drops. It often comes with other low-frequency components (fig.\ref{fig_spectra}.) We compared the results obtained by processing the data by both linear and non-linear models with no harmonics included to the results obtained by including one harmonic with known frequency and varying the frequency in the range of (16--24)~Hz with~1 Hz increment in every processing run.

The inclusion of the harmonic into either model did not reduce the scatter of the gravity estimate. The variation of the frequency was changing the result of the linear estimate in the range of $\pm$6~\muG, while the non-linear estimate was changing up to $\pm$21 \muG. The experiment has confirmed better stability of the linear model, but it also has shown that the low-frequency disturbances of the IMGC-02 gravimeter can not be successfully addressed by including harmonic components into either linear or non-linear model. We did further comparison of the models with no harmonic components included.

\begin{figure}[ht]
\centering
\includegraphics[height=150mm]{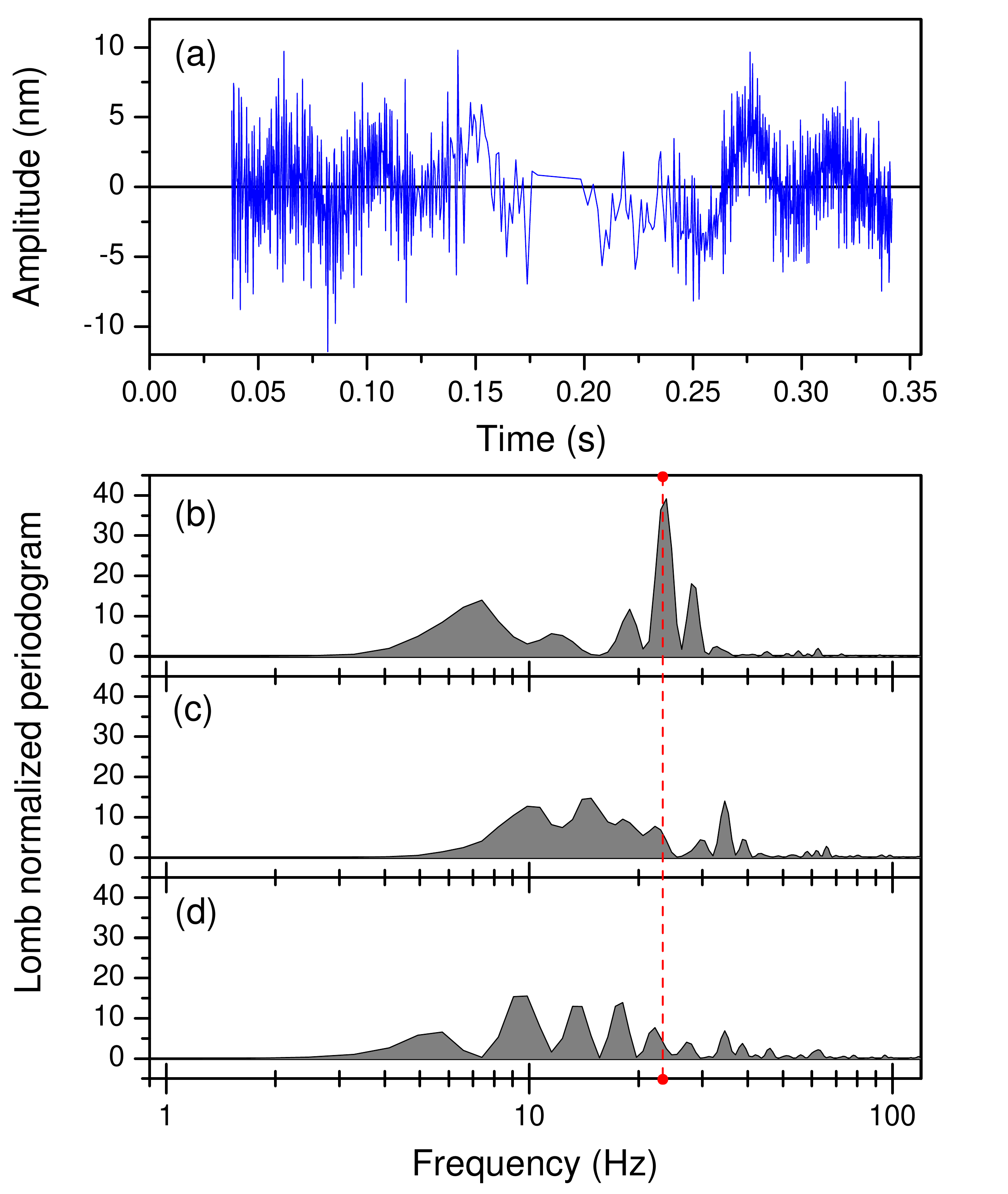}
  \caption[short title]
  {
Typical single drop residuals (a) and the normalized Lomb-Scargle periodogram for three consecutive drops (b), (c), (d). Dashed line corresponds to the frequency of 22~Hz.
 }
\label{fig_spectra}
\end{figure}

\section{Results}
\label{sec:results}
We compared linear and non-linear models by applying them to the five data sets taken at the following sites during the period of 2011-2013:
\begin{itemize}
\item Turin (IT): gravimetric laboratory in the basement of the INRIM building;
\item Walferdange (LU): gravimetric laboratory in the old mine;
\item Frejus (FR): underground particle physics laboratory;
\item Cluj-Napoca (RO): open industrial site used for pressure measurements;
\item Cosenza (IT): gravimetric laboratory in the basement of the University building.
\end{itemize}
In comparing the models we did not apply any instrumental or environmental corrections that affect both models in the same way. The only exception was the tidal corrections, as tides strongly affect the normality of the observations required for the three sigma rule, the only rejection applied to the data.
Tables \ref{tab_NL}  and \ref{tab_L} show parameters obtained for the same drop by non-linear and linear models, the table \ref{tab:dataset} compares the results for different datasets.
\begin{table}[ht]
\caption{Results of non-linear drop processing}
\begin{indented}
\item[]\begin{tabular}{@{}cccc}
\br
Estimated quantity & Value & Standard deviation & Unit\\
\mr
$g_a$    & 980 533 954.0         & 15.5                  & \muG      \\
$\gamma$ & 9.2 $\times 10^{-6}$  & 7.9  $\times 10^{-6}$ & s$^{-2}$   \\
$h_b$    & 0.01895                & $< 10^{-4}$           & m  \\
$g_b$    & 980 534 079.6         & 3.9                   & \muG  \\
$\phi$   & 6.0 $\times 10^{-7}$  & 2.3  $\times 10^{-7}$ &  s$^{-1}$  \\
$t_a$    & 0.1520                & $< 10^{-5}$           & s  \\
$z_a$    & 0.1133                & $< 10^{-5}$           & m  \\
$d$      & -136415.2            & 1.3                    & nm  \\
\br
\end{tabular}
\end{indented}
\label{tab_NL}
\end{table}

\begin{table}[ht]
\caption{Results of linear drop processing}
\begin{indented}
\item[]\begin{tabular}{@{}ccccc}
\br
Estimated quantity & Value & Formula/Notes &  Unit\\
\mr
$g$            &  980 534 078.5   & (\ref{eq_LSS})  & \muG   \\
$V_0$          & 1.866    & (\ref{eq_LSS})  &  m s$^{-1}$  \\
$S_0$          & -0.06447   & (\ref{eq_LSS})  &  m  \\
$h_{\rm eff}$  &  0.01895   & (\ref{eq_heff}) &  m  \\
$g \uparrow $  &  980 534 098.2   & (\ref{eq_LSS}) applied to left branch   & \muG   \\
$g \downarrow$ &  980 534 033.6   &  (\ref{eq_LSS}) applied to right branch & \muG   \\
$t_a$          &  0.1903   & apex of last iteration, see \ref{sec_apex_step} &  s  \\
$\D w_g(t_a) / \D t$&  -0.2193  & (\ref{eq_dS1}) substituted to (\ref{eq_GTS_lin_comb})    &   s$^{-1}$  \\
$d$            &  9.51           &  (\ref{eq_res_step})  & nm \\
$T$            &  0.3039   &  $T_N - T_1$   &  s  \\
$H$            &  0.1132   & $g\,T^2/8$   &  m  \\
$C_1$          &   7.153 $\times 10^{-16}$ & (\ref{eq_C1_central}) &  s  \\
$\phi$         &  -3.788 $\times 10^{-7}$  & (\ref{eq_phi}) &  s$^{-1}$  \\
$\Delta g_d$   &  -0.208   & (\ref{eq_Delta_gd})  & \muG   \\
$\Delta \overline{g}_{\phi}$ & -2.656 $\times 10^{-13}$  & (\ref{eq_dg_phi}) & \muG   \\
\br
\end{tabular}
\end{indented}
\label{tab_L}
\end{table}
\begin{table}[ht]\small
\caption{\it Comparison of results obtained with nonlinear (NL) and linear (L) models }
\begin{tabular}{@{}ccccccc}
\br
Site                 &  model & rejected& $h_{\rm ref}$ & $\bar{g}$     &  $\bar{\sigma}$ & $\bar{g}_{_{\rm{L}}} - \bar{g}_{_{\rm{NL}}}$ \\
                     &        &  \%     & m         & \muG          &  \muG & \muG \\
\mr
Turin, IT             &  NL    & 1.0     & 0.4758    & 980 534 193.6 &  18.5 &  -0.3 $\pm$ 0.8  \\
2011-11-10, $N$=1000 &  L     & 1.2     & 0.4753    & 980 534 193.3 &  18.1 &                  \\
                     &        &         &           &               &       &                  \\
Walferdange, LU       &  NL    & 0.8     & 0.4747    & 980 964 165.8 &  22.5 &  -0.3 $\pm$ 0.8  \\
2011-11-03, $N$=1400 &  L     & 0.7     & 0.4746    & 980 964 165.5 &  22.0 &                  \\
                     &        &         &           &               &       &                  \\
Frejus, FR            &  NL    & 2.7     & 0.4752    & 980 095 592.8 &  25.1 &  -0.3 $\pm$ 1.0  \\
2013-10-12, $N$=1200 &  L     & 4.0     & 0.4751    & 980 095 592.5 &  23.4 &                  \\
                     &        &         &           &               &       &                  \\
Cluj-Napoca, RO       &  NL    & 1.4     & 0.4781    & 980 689 520.0 &  20.5 &  1.8 $\pm$ 0.9   \\
2013-04-17, $N$=1000 &  L     & 1.1     & 0.4779    & 980 689 521.8 &  20.3 &                  \\
                     &        &         &           &               &       &                  \\
Cosenza, IT           &  NL    & 1.0     & 0.4739    & 980 106 548.2 &  21.2 &  0.6 $\pm$ 0.9   \\
2013-10-05, $N$=1000 &  L     & 3.0     & 0.4739    & 980 106 548.8 &  20.1 &                  \\
\br
\end{tabular}
\label{tab:dataset}
\end{table}

The values reported for the linear and non-linear models correspond to the reference heights found as
\begin{equation}
\label{eq_h_ref_LM}
h_{\rm ref \, L} = h_{\rm apex} - h_{\rm eff},
\end{equation}
\begin{equation}
\label{eq_h_ref_NLM}
h_{\rm ref \, NL} = h_{\rm apex} - h_b,
\end{equation}
where $h_{\rm apex}$ is the distance from the site mark to the apex of the trajectory found as (fig.~\ref{fig_heights})
\begin{equation}
\label{eq_h_apex}
h_{\rm apex} = h_{\rm start} + h_{\rm cut} + H.
\end{equation}
Here $h_{\rm start} \approx 329$~mm is the distance from the site mark to the resting position of the test mass, $h_{\rm cut}\approx 51$~mm is the distance from the start of the motion to the first level adopted for processing (the values may vary from site to site), $H$ is the height of the upper part of the trajectory used to derive $g$.
\begin{figure}[t]
\centering
\small
\includegraphics[height=100mm]{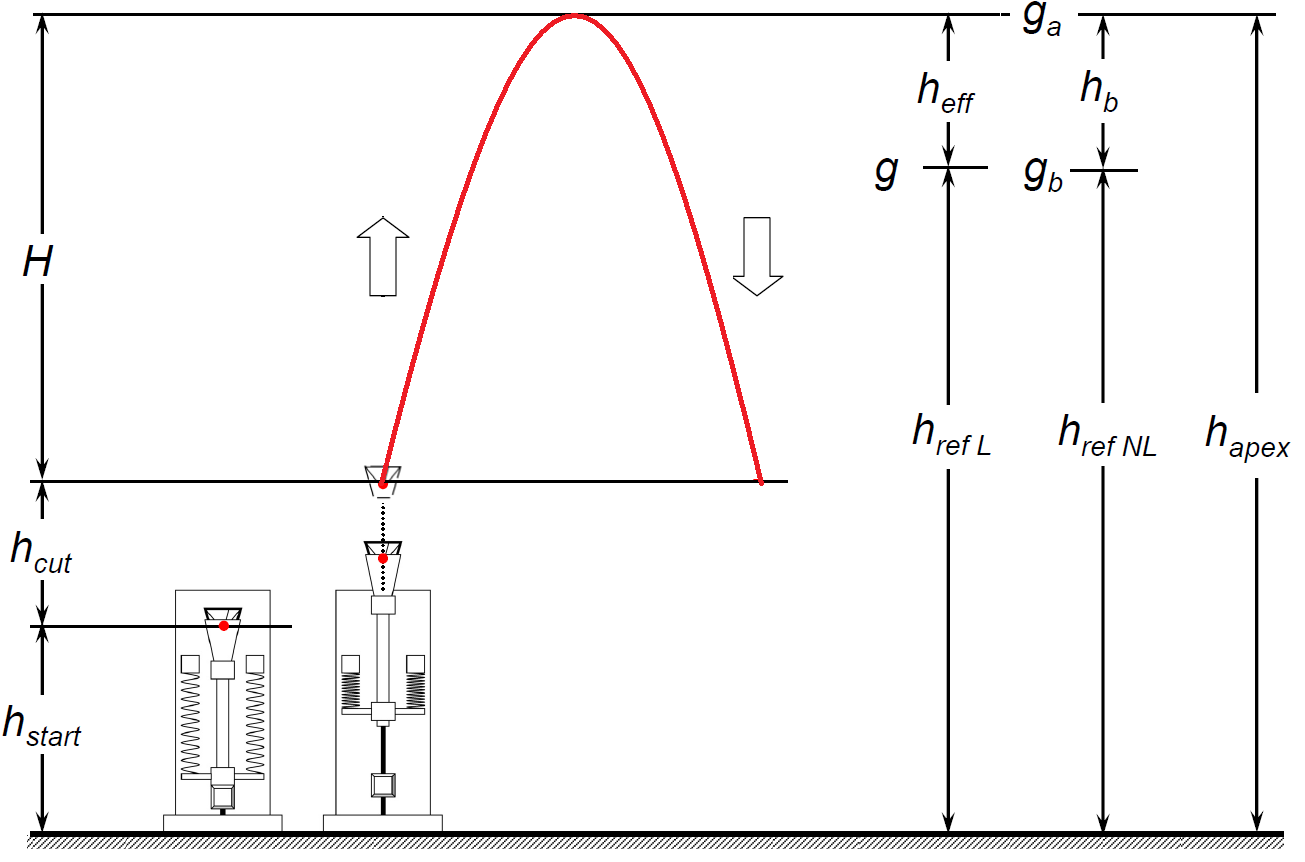}
  \caption[short title]
  {
Evaluation of reference heights for the linear and non-linear models.
  }
\label{fig_heights}
\end{figure}
For all the datasets the results obtained by both models (tab.~\ref{tab:dataset}) are in good agreement. The maximum observed bias of the non-linear model reaches 1.8~$\pm$~0.9~\muG. This value is within the uncertainty budget of the instrument, but is not explained by the difference in reference heights that reaches only 0.2 mm.

\section{Conclusions}\label{sec:conclusion}
The non-linear regression model has been used in the IMGC-02 gravimeter to resolve the singularity at the apex of the trajectory caused by the discrete fringe counting in the Michelson-type interferometer. We have developed and implemented the new measurement procedure based on the linear regression models. The issues of the rise-and-fall trajectory have been addressed by the procedure as following.
\begin{itemize}
\item {\emph {Apex step}.} We developed a precess that iteratively reduces the apex step till its influence on the measured gravity is below 0.25 \muG of random error in one drop.
\item {\emph {Verical gravity gradient}.} The constant vertical gradient is accounted by relating the result to the effective measurement height, at which the actual gravity equals the measured one regardless of the magnitude or the uncertainty of the gradient.
\item {\emph {Velocity-proportional disturbances}.} The linear model is insensitive to the velocity-proportional disturbances, as they cancel out on the symmetric trajectory. The linear model allows to monitor the symmetry of the trajectory and correct its imbalance, if necessary.
\item {\emph {Harmonic disturbances}.} The laser modulation creates random error of 0.3~\muG or less in one drop, and so is insignificant. The seismometer resonance of 22 Hz can cause systematic error of up to 10~\muG and remains the major problem of the IMGC-02 gravimeter. The problem can not be addressed by including this component into either linear of non-linear model. Future efforts may include correcting the result by independently measured accelerations of the reference reflector~\cite{araya2013}.
\item {\emph {Other disturbances of the trajectory}.} The corrections for other disturbances, if necessary, can be derived and applied to the measurement results. Due to the linearity of the model, the new corrections do not interfere with the existing ones and can be applied to the existing results without re-processing the data.
\end{itemize}
While the bias is an intrinsic property of any non-linear model, the theoretical evaluation of the bias is a difficult task with no guaranteed solution. 
For the first time we were able to evaluate the bias by the direct drop-by-drop comparison of the non-linear and linear models. The maximum observed bias was about 2~\muG at the site with the worst measurement conditions. 
This is very optimistic result that confirms validity of the earlier measurements that involved non-linear fitting and translation of the result to the best reference height, as discussed in chapter \ref{sec:NLmodel}. 
In addition, the new data processing method developed for the IMGC-02 instrument and based on the linear model of the trajectory provides the following advantages
\begin{itemize}
\item Clear and translucent measurement procedure based on rigorous theory providing unbiased estimates of the gravity parameter traceable to the base units.
\item Independent analysis of each disturbance, deriving of the corrections that can be retrospectively applied to the existing measurements without re-processing the original data.
\item Possibility to correct the disturbances of the reference reflector by its independently measured accelerations.
\end{itemize}
Because of the advantages of the linear model discussed in this work, we strongly recommend that users of rise-and-fall gravimeters consider employing the model in their instruments.
\section*{Acknowledgment}
The Istituto Nazionale di Ricerca Metrologica (INRiM) gave the authors the possibility to work with the gravimeter IMGC-02 and to use all the data collected in the last two years. The work of Sergiy Svitlov was partially supported by the Deutsche Forschungsgemeinschaft (DFG, Germany) under the project SV 86/1-1.


\section*{Appendix}
\label{sec:appendix}
To find the analytic expression for the limiting shapes of the weighting functions, we start with the formula (\ref{eq_ws}) with substituted coefficients $a_i$ (\ref{eq_GTS_ai})
\begin{equation}
\label{eq_ws_exp}
\begin{array}{ll}
w_s(t) & =  \sum a_i \, \delta(t-T_i)
\\
& = 2\left|
\begin{array}{ccc}
\sum T_{i}^0  & \sum T_{i} & \sum \delta(t-T_i) \\
\sum T_{i} & \sum T_{i} ^{2} & \sum T_{i} \delta(t-T_i) \\
\sum T_{i}^{2} & \sum T_{i}^{3} &  \sum T_i^2 \delta(t-T_i)
\end{array}
\right| :\left|
\begin{array}{ccc}
\sum T_{i}^0 & \sum T_{i} & \sum T_{i} ^{2} \\
\sum T_{i} & \sum T_{i} ^{2} & \sum T_{i} ^{3} \\
\sum T_{i}^{2} & \sum T_{i}^{3} & \sum T_{i}^{4}
\end{array}
\right|.
\end{array}
\end{equation}
We then multiply each sum by $\Delta z$, the distance between the neighbouring levels. This does not change the value of (\ref{eq_ws_exp}), as both determinants get the same multiplier $(\Delta z)^3$ -- one $\Delta z$ per column. Now (\ref{eq_ws_exp}) has two types of sums:
\begin{equation}
\label{eq_sums}
\sum T_{i}^n \Delta z 
\;\; \textrm{and} \;\;
\sum T_i^m \delta(t-T_i)\Delta z,
\end{equation}
where $n$ runs from 0 to 4, $m$ runs from 0 to 2. The distance $\Delta z$ is the difference of the coordinates of the consecutive levels, however the order of the coordinates is different on the upward and the   downward brances, i.e.
\begin{equation}
\label{eq_Dz_up}
\Delta z \uparrow \; = z(T_{i+1})-z(T_i) ,
\end{equation}
\begin{equation}
\label{eq_Dz_down}
\Delta z \downarrow \; = z(T_i) - z(T_{i+1}).
\end{equation}
Both cases can be combined as
\begin{equation}
\label{eq_Dz_module}
\Delta z = |z(T_{i+1})-z(T_i)|, 
\end{equation}
so the sums can be rewritten as
\begin{equation}
\label{eq_sums_RS}
\sum T_{i}^n |z(T_{i+1})-z(T_i)| \;\; \textrm{and} \;\;
\sum T_i^m \delta(t-T_i)|z(T_{i+1})-z(T_i)|.
\end{equation}
As the number of levels increases, the sums (\ref{eq_sums_RS}) turn into the Riemann-Stiltjes integrals
\begin{equation}
\label{eq_int_RS}
\int_{-T/2}^{T/2} \tau^n \D |z(\tau)| \;\; \textrm{and} \;\;
\int_{-T/2}^{T/2} \tau^m \delta(t-\tau) \D |z(\tau)|.
\end{equation}
Substitution
\begin{equation}
\label{eq_dz}
\D |z(\tau)| = g|\tau|\D \tau
\end{equation}
turns (\ref{eq_int_RS}) into the straight Riemann integrals 
\begin{equation}
\label{eq_int_RS}
\int_{-T/2}^{T/2} \tau^n |\tau| \D \tau 
\;\; \textrm{and} \;\;
\int_{-T/2}^{T/2} \tau^m |\tau| \delta(t-\tau) \D \tau.
\end{equation}
Due to the symmetry, the first integral equals zero for the odd $n$'s. For the even $n$'s the integral is
\begin{equation}
\label{eq_1_int}
\int_{-T/2}^{T/2} \tau^n |\tau| \D \tau 
=
\frac{2}{n+2}\left(\frac{T}{2}\right)^{n+2}
\end{equation}
The second integral, due to the sampling property of the $\delta$-function (\ref{eq_Si_via_ddelta}), is 
\begin{equation}
\int_{-T/2}^{T/2} \tau^m |\tau| \delta(t-\tau) \D \tau = (-t)^m|-t|.
\end{equation}
By substituting these values into (\ref{eq_ws_exp}) and expanding the determinants, we get
\begin{equation}
\label{eq_ws_inf}
\begin{array}{ll}
w_s(t)_{_{N \rightarrow \infty}} & = 2\left|
\begin{array}{ccr}
\frac{T^2}{4} & 0 & |t| \\
0 & \frac{T^4}{32}  & -t|t|\\
\frac{T^4}{32}  & 0 & t^2|t| 
\end{array}
\right| :\left|
\begin{array}{ccc}
\frac{T^2}{4} & 0 & \frac{T^4}{32} \\
0 & \frac{T^4}{32}  & 0 \\
\frac{T^4}{32}  & 0 & \frac{T^6}{192}
\end{array}
\right|
= \frac{1536}{T^6} t^2 |t| - \frac{192}{T^4} |t|.
\end{array}
\end{equation}
The asymptotic weighting functions by velocity and acceleration are found by integration (\ref{eq_ws_wv_wg}):
\begin{equation}
\label{eq_wv_inf}
w_{_V}(t)_{_{N \rightarrow \infty}} = 
- \int \left(
\frac{1536}{T^6} t^2 |t| - \frac{192}{T^4} |t|
\right) \D t = 
-\frac{384}{T^6} t^3|t| + \frac{96}{T^4} t|t|,
\end{equation}
\begin{equation}
\label{eq_wg_inf}
w_g(t)_{_{N \rightarrow \infty}} = 
- \int \left(
-\frac{384}{T^6} t^3|t| + \frac{96}{T^4} t|t|
\right) \D t = 
\frac{76.8}{T^6} t^4|t| - \frac{32}{T^4} t^2|t| + \frac{1.6}{T}.
\end{equation}
The constants of integration were determined based on the conditions (\ref{eq_wz_wv_wg_square}).
The asymptotic weighting functions (\ref{eq_ws_inf}, \ref{eq_wv_inf}, \ref{eq_wg_inf}) are valid for any rise-and-fall absolute gravimeter with levels equally spaced in distance and $g$ found by the LS-adjustment of the model (\ref{eq_3_param}).
If the levels are equally spaced in time, the sums of (\ref{eq_ws_exp}) can be multiplied by $\Delta t$ (the time inteerval between levels), leading directly to the Riemann integrals as $N \rightarrow \infty$ and producing the following weighting functions:
\begin{equation}
\label{eq_ws_inf_EST}
w_s(t)_{_{N \rightarrow \infty}} = 
 \frac{360}{T^5} t^2 - \frac{30}{T^3} ,
\end{equation}
\begin{equation}
\label{eq_wv_inf_EST}
w_{_V}(t)_{_{N \rightarrow \infty}} = -\frac{120}{T^5} t^3 + \frac{30}{T^3} t,
\end{equation}
\begin{equation}
\label{eq_wg_inf_EST}
w_g(t)_{_{N \rightarrow \infty}} = 
\frac{30}{T^5} t^4 - \frac{15}{T^3} t^2 + \frac{1.875}{T}.
\end{equation}
If the data from only one branch of the parabola are used for the fitting, the limits of integration in (\ref{eq_int_RS}) will change accordingly. The weighting functions for the downward branch will in fact be the same as for the free-fall gravimeters \cite{nagornyi1995}. 

The method considered here is an alternative to the one described in \cite{nagornyi1995}, where the double integration (\ref{eq_wv_inf}, \ref{eq_wg_inf}) is applied at the beginning rather than at the end.

\newpage

\section*{References}


\end{document}